\RequirePackage{fix-cm}
\documentclass[smallextended]{svjour3}       
\smartqed  
\usepackage{graphicx}	
\usepackage{hvfloat,lipsum}
\begin{document}

\title{A Systematic Mapping Study on Cloud Computing
}


\author{Jose Fernando S. Carvalho \and
          Paulo Anselmo da Mota Silveira Neto \and Vin'cius Cardoso Garcia \and Rodrigo Elia Assad \and Frederico Durao
}


\institute{Jose Fernando S. Carvalho  \at 
 Informatic Center, Federal University of Pernambuco, Recife, PE, Brazil. \\
	 \email{jfsc@cin.ufpe.br}
           \and
           Vinicius Cardoso Garcia \at
            \email{vcg@cin.ufpe.br}
            \and
            Paulo Anselmo da Mota Silveira Neto \at
            \email{pamsn@cin.ufpe.br} 
            \and
            Rodrigo Elia Assad \at
            \email{assad@deinfo.br} 
            \and
            Frederico Durao \at
            \email{freddurao@dcc.ufba.br} 
}

\date{Received: date / Accepted: date}

\maketitle

\begin{abstract}
Cloud Computing emerges from the global economic crisis as an option to use computing resources from a more rational point of view. In other words, a cheaper way to have IT resources. However, issues as security and privacy, SLA (Service Layer Agreement), resource sharing, and billing has left open questions about the real gains of that model.  This study aims to investigate state-of-the-art in Cloud Computing, identify gaps, challenges, synthesize available evidences both its use and development,  and provides relevant information, clarifying open questions and common discussed issues about that model through literature. The good practices of systematic mapping study methodology were adopted in order to reach those objectives. Although Cloud Computing is based on a business model with over 50 years of existence, evidences found in this study indicate that Cloud Computing still presents limitations that prevent the full use of the proposal on-demand.

\end{abstract}

\section{Introduction}
\label{intro}

Since 60's, researchers as Douglas Parkhill and John McCarthy, has been spending efforts on development of a computing model named Utility Computing. That model should enable use of computing resources, in the same way of Electricity and Gas. In other words, computing on-demand \cite{fish:McCarthy} \cite{fish:Parkhill}.

The adoption of the model showed positive responses from the market. However, their weakness was exposed during the oil crisis, 70 years, and then the emergence of PCs, 80 years, when the model enters into disuse \cite{fish:Martin}.

Past thirty years, the world is faced with the current global economic crisis. Consequently, the search for cost reduction is intensified. In this scenario, the Utility Computing model emerges again. However, with a new name, Cloud Computing \cite{fish:Chieu Trieu}.

Nowadays, Cloud Computing is considered a model for enabling convenient, on-demand network access to a shared pool of configurable computing resources that can be rapidly provisioned and released with minimal management effort or service provider interaction. By promoting greater flexibility and availability at lower cost, that model has been receiving a good deal of attention lately  \cite{fish:WayneA}.

However, the attention directed to the model also comes with open issues that confront the real efficiency of Cloud Computing among which the conception of contracts of service \cite{fish:Kertesz2}\cite{fish:Nae}\cite{fish:Luo}, real economic benefits \cite{fish:Kondo}, choice of a suitable software architecture to SaaS developing  \cite{fish:Chapman}\cite{fish:Kossmann}, data privacy \cite{fish:WayneA} \cite{fish:Zhou}, adoption of agile process \cite{fish:Guha}, and LAWs \cite{fish:Taylor} \cite{fish:Doelitzscher} \cite{fish:ChoBrian} can be cited.

In order to investigates and promote advances on the Cloud Computing, this study aims to provide relevant information to community, clarifying and  synthesizing available evidences to suggest important implications for practice, as well as, identifying research trends and new gaps. 

Thus, good practices from Systematic Mapping Studies (MS) \cite{fish:KPetersen} and Systematic Reviews (SR) \cite{fish:BKitchenham}  were combined to ``map out'' the Cloud Computing through the main question: \textbf{what are the main problems and solutions in Cloud Computing?} 

The remainder of the paper is structured as follows: Section II describes the research method used. Section III presents the main findings, Section IV  with analysis of the results and the mapping of studies, Section V presents a discussion, Section VI presents the threats to validity, and Þnally, Section VII discusses the main conclusions.

\section{Research Methodology}\label{method}
A Systematic Review is way of identifying, evaluating, interpreting and comparing all available researches which are relevant to a particular question \cite{fish:BKitchenham}. Thus, that process produces detailed answers to specific scope. On the other hand, a Systematic Mapping Study uses a more quantitative than qualitative approach, which intends to ``map out'' the undertaken research \cite{fish:KPetersen}.
\begin{figure*}[!t]
\centering
\includegraphics[width=4.6in]{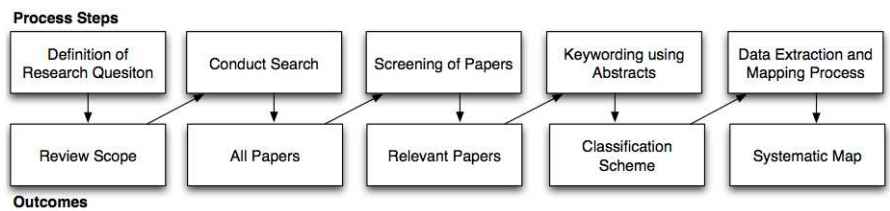}
\caption{The Systematic Mapping Process \cite{fish:KPetersen}.}
\label{fig_processMS}
\end{figure*}

\subsection{Research Questions}
Eight sub-questions were derived from fundamental question:
\begin{itemize}
\item	\textbf{RQ1. Which challenges were found regarding to economic problems?}
\newline
It aims to identify factors influence the adoption of Cloud computing, Grid Computing and challenges faced by Service Providers (Cloud Providers) and Customers.

\item \textbf{RQ2. What problems and solutions were found regarding to SLA?}
\newline
The objective of this question is to understand the role of SLA in Cloud Computing, identifying its challenges and used techniques to ensure QoS (Quality-of-Service) in the environment.

\item	\textbf{RQ3. What are the Cloud Computing social impact?}
\newline
This question seeks to address topics refer to possible Government Cloud Computing usage, conflicts in laws, and impacts over citizens.
\item	\textbf{RQ4. What are the challenges found regarding to service conception on Cloud Computing environments?}
\newline
This question to aims to identify requirements and challenges for the development of infrastructure and datacenter software (SaaS).
\item	\textbf{RQ5. What are the main challenges regarding to the Elastic property?}
\newline
At this point, the purpose is to address techniques refer to resource allocation on Cloud computing.

\item	\textbf{RQ6. What are the problems and solutions about data storage?}
\newline
The objective of this question is to address both the problems and techniques used to solve them.

\item	\textbf{RQ7. How is performed the resource usage monitoring on Cloud Computing?}
\newline
It aims to understand how is made the resource usage monitoring in Cloud Computing, and obtain tools and techniques.

\item	\textbf{RQ8. Which are the main security challenges?}
\newline
A overview about Security challenges on Cloud Computing.

\end{itemize}

\subsection{Search Strategy, Data Sources and Study Selection}
\begin{figure}[!t]
\centering
\includegraphics[width=3.5in]{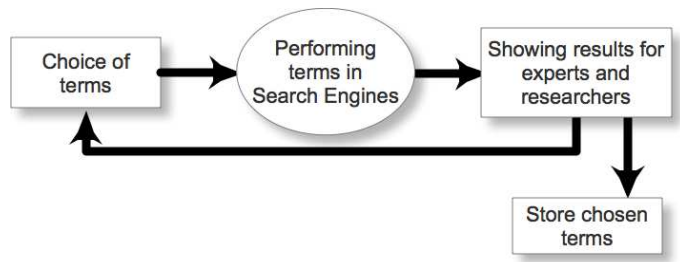}
\caption{Cycle for refine Terms.}
\label{fig_cycleOfTerms}
\end{figure}

Based on approach described in \cite{fish:BKitchenham3}, an initial set of terms were chosen in order to answer the eight Research Questions. In turn, those terms was combined and tested in search engines. Then, the results are showed to experts and researchers of Social Machines Research Team\footnote{https://sites.google.com/site/socialmacslab/}, which the author is member,  in order to refine the terms.
Case, the results were satisfactory, the terms were chosen. Otherwise, the cycle would begin again, as follow on Figure ~\ref{fig_cycleOfTerms}.

The result of refining of terms enabled the conception of a complete list of search strings and their combination as follow on Table \ref{table_Strings}. It is important to emphasize that these strings were calibrated in order to deal with possible problems regarding to the use of different search engines.
\begin{table}[hbp]
\renewcommand{\arraystretch}{1.3}
\caption{Search Strings.}
\label{table_Strings}
\centering
\begin{tabular}{p{11cm}}
\hline
\\
 \bfseries \hspace{27 mm} ("cloud computing" OR Cloud)\\
\\
\bfseries \hspace{47 mm}AND \\
\\
\bfseries (business OR challenge OR problem OR market  OR outsourcing 
\bfseries OR resource OR management OR elastic OR provisioning OR 
\bfseries control OR "SLA" OR QoS OR data OR "Utility computing"  OR 
\bfseries "Grid Computing" OR Security OR Vulnerability OR Knowledge 
\bfseries OR Government OR science OR storage OR Service OR 
\bfseries monitoring OR "Open source" OR tool OR Virtualization OR role 
\bfseries OR medical OR green OR Protection OR laws OR acts OR privacy 
\bfseries OR health OR Architecture OR model OR saas OR software OR 
\bfseries application)\\
\\
\hline
\end{tabular}
\end{table}

We considered publications retrieved from: ACM Digital library, ScienceDirect, IEEE Xplore, EL COMPENDEX, SCOPUS, and DBLP. Since it was not possible to have syntactically identical search strings for all the searched databases, all of the search strings were logically and systematically checked by more than one author. In addition, two further levels of search were performed, a manual search was performed in thirty-four different conferences and fourteen journals related to the topic addressed by the study.

The studies were included if they involved:

\begin{itemize}
\item	Research that explores Cloud Computing as mainly focus;
\item	Studies that address utility computing linked to Cloud Computing;
\item	Studies that address comparison among Cloud and Grid Computing;
\end{itemize}

The latter, excludes the studies which: 

\begin{itemize}
\item	Studies that donÕt have Cloud as the main focus;
\item	Duplicated Studies;
\item	Keynotes, Presentations and Whitepapers.
\end{itemize}

\begin{figure}[!t]
\centering
\includegraphics[width=3.0in]{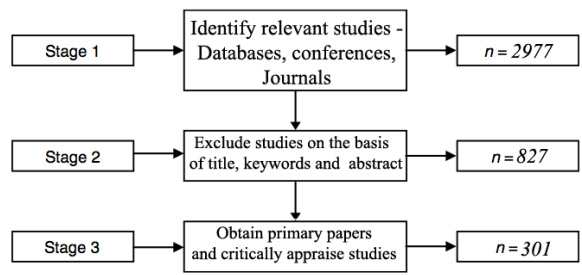}
\caption{Stages of the selection process.}
\label{fig_selectionStage}
\end{figure}

In the initial stage were found 2977 studies. The inclusion and exclusion criteria were applied on the title, keywords, and abstract of the identified studies, resulting in 827 studies. In that, a deep reading was performed on abstracts. If they clearly not present its proposal, a deep reading was performed on all content. Then, the studies were excluded or not. This way, resulting in 301 documents.

\subsection{Classification Scheme}
The key wording process is a way to ensuring that the scheme  takes the existing studies into account \cite{fish:KPetersen}. Besides use key wording process, all content of studies were read in order to obtain its classification.

\begin{table}[hbp]
\renewcommand{\arraystretch}{1.3}
\caption{Cloud aspects based Facet.}
\label{table_AspectBasedFacet}
\centering
\begin{tabular}{|l| p{8cm}|}
\hline
\bfseries Issue/Aspect & \bfseries Description\\
\hline 
Resource Management & Studies that have developed your content under the influence of aspects related to flexibility of the Cloud.\\
\hline
Architectural &Studies that are concerned with structural aspects of the cloud showing the care with this point in its context.\\
\hline
Economic & Refer to economic terms in content. In this 
point we refer to investment, business model, financial data and costs.\\
\hline
Security & Refer aspects as privacy, vulnerabilities 
or information security.\\
\hline
\end{tabular}
\end{table}

In this work were used 2 set of facet, one facet structured the topic in terms of the research questions. The other, considered important issues discussed in Cloud area \cite{fish:WayneA} \cite{fish:Tao Q} \cite{fish:Gagliardi} \cite{fish:LinShih} and are detailed on Table ~\ref{table_AspectBasedFacet}.

\section{Main Findings}
In this section, each topic presents the findings of a research question, highlighting evidences gathered from data extraction process. These results populate the classification scheme, which evolves while doing the data extraction.

\subsection{RQ1 - Which challenges were found regarding to economic problems?}
The current economic conditions and the need to lower costs have forced enterprises to consider the adoption of Cloud Computing \cite{fish:Chieu Trieu}. This is because the promise of Cloud Computing is to deliver all the functionality of existing IT services (and in fact enable new functionalities that are hitherto infeasible) even as it dramatically reduces the upfront costs of computing that deter many organizations from deploying many cutting-edge IT services \cite{fish:Marston}.

In addition to lower upfront costs, organizations can to see your substantial investments grossly underutilized. This is because this servers are using among 10-30\% of itself available computing power even as desktop computers have an average capacity utilization of less than 5\% \cite{fish:Marston}. However, economics problems do not stay on enterprises only, but extends to science field too. In that case were found issues about maintainability of scientific projects and prediction of costs in beginning of the projects \cite{fish:Kondo}\cite{fish:Hou_Zhengxiong}.

For Marston et al. \cite{fish:Marston}, Cloud Computing enables the competitive market for small businesses. The reason for this observation refers to investment that can be done gradually while using the Cloud \cite{fish:Kondo}.

However, while Cloud environments attract stakeholders through low cost promise, it is important to note that there are technological alternatives that may be more interesting due to the initial stage of the Cloud. In some cases, a Grid Computing may be more appropriate. Given its maturity and existing solutions (caBIG\footnote{http://cabig.cancer.gov/}, Earth System Grid\footnote{http://www.earthsystemgrid.org/home.htm}).
\begin{table}[hbp]
\renewcommand{\arraystretch}{1.3}
\caption{Grid vs. Cloud \cite{fish:Weinhardt}}
\label{table_GridVsCloud}
\centering
\begin{tabular}{| p{3.5cm}| p{3.0cm}| p{3.5cm}|  }
\hline
\bfseries Criteria & \bfseries Grid Computing  & \bfseries Cloud computing \\
\hline
Virtualization & in its Beginning & essential\\
\hline
Type of Appliaction & batch & interactive\\
\hline
Development of applications & local & In the Cloud\\
\hline 
\end{tabular}
\end{table}

\begin{table}[!t]
\renewcommand{\arraystretch}{1.3}
\centering
\begin{tabular}{| p{3.5cm}| p{3.0cm}| p{3.5cm}|  }

\hline 
Access & via Grid middleware & via standard Web protocols\\
\hline
Organizations & virtual & physical\\
\hline
Business Model & sharing & pricing(utility model)\\
\hline
Control & decentralized & Centralized (data center)\\
\hline
Switching Cost & low due to standardization & high due to incompatibilities\\
\hline
SLAs/Liability &  not yet enforceable & essential\\
\hline
Openness &  high & low\\
\hline
\end{tabular}
\end{table}

The choice between Cloud or Grid Computing as a new platform must be a careful task. We found some properties that differ Cloud Computing of Grid Computing \cite{fish:Weinhardt}. Those properties are showed on Table \ref{table_GridVsCloud}.

According to \cite{fish:Kondo}, in compare with Grid Computing, due to control enabled by Cloud Computing, a scientific project based on VC (Volunteer Computer) can reach 40-95\% saving costs, when deployed in Cloud. In this case, Amazon WS.  

But, the numbers claims that Cloud Computing have an expensive cost in concern with bandwidth usage and the savings occur when the system works with less than 10TB in storage. Otherwise, Cloud Computing becomes less viable. According to the author, the use of the band was the main factor. Also, the results of that work are available\footnote{http://mescal.imag.fr/membres/derrick.kondo/cloud\_calc.xlsx} for future analysis.
\begin{figure}[hbp]
\centering
\includegraphics[width=2.5in]{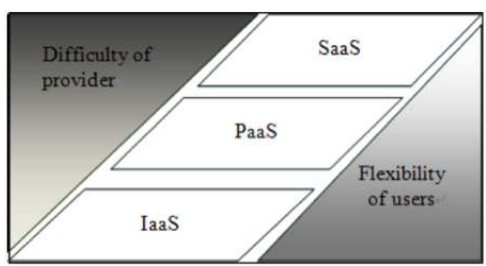}
\caption{Flexibility of Cloud deployment Models \cite{fish:Shi A}.}
\label{fig_flexibilityOfModels}
\end{figure}
The flexibility enabled by Cloud deployment models, also presents problem. While the customer requires major flexibility of a particular service from a Cloud Provider, enabling this flexibility can be a difficult task, depending of adopted service model  (SaaS, PaaS or IaaS). The Figure \ref{fig_flexibilityOfModels} represents the scenario well.

For Shi et al. \cite{fish:Shi A}, from a Cloud Provider perspective, it is more viable customize an IaaS service because this model provides resources as storage space, processing and bandwidth. On the other hand,  SaaS customizing becomes a difficult task, because the requirements for a customer may not be feasible to implement in that software.

It is important to remember that choosing a certain degree of customization in a given service model (SaaS, PaaS or IaaS) should be linked directly with the form of charge for the service. Otherwise, the provider can not obtain the desired profit for its solution.

According to Schuff and Altaf \cite{fish:Schuff}, due to lack of techniques to adapt the service, Cloud Providers remove features and other elements of the application in order to reduce costs and enable that SMEs (small and medium enterprises) have access to that service. That practice, too, becomes a barrier to adoption of solutions in Cloud Computing.

In context of Private Clouds, the choice of a right technology can be a great ally to reach costs reduction on Cloud environment. In \cite{fish:ChoBrian}, the Pandora Planning System (People and Networks Moving Data Around) represents an effort to minimize costs over the sent of large datasets among remote servers, providing a way to automatically opt between to send this data over Internet or Shipping in physical HardDisks.

Thus, if Pandora algorithms point for to send the data in a HardDisk, the system communicates with a Shipping service (in this case the FedEX) and sends it. Otherwise, the data will be sent via the Internet.

\subsection{RQ2 - What problems and solutions were found regarding to SLA?}\label{sla}

SLA (Service Layer Agreement) is a formal negotiated agreement between Service Provider (Cloud Provider) and Customer and its terms refer to quality and responsibilities from each part. When used appropriately, the SLA should be \cite{fish:Kandukuri}:
\begin{itemize}
\item Identify and define customer needs;
\item	Provide a model for understanding;
\item	Simplifying complex issues;
\item	Reduce conflicts;
\item Encourage dialogue on any disputes;
\item Eliminate unrealistic expectations.
\end{itemize}

Through the terms of the SLA, Cloud Providers declare its level of Quality-of-Service (QoS) according to its \textbf{actual capability} \cite{fish:Macias 2}. This way QoS has been an important factor in choose process of a Cloud Provider \cite{fish:Tao Q}.

In this context, this study identified considerable attention among the leading authors about  to keep the QoS at an acceptable level  \cite{fish:Nae}\cite{fish:Luo}\cite{fish:Kandukuri}\cite{fish:Chaves}\cite{fish:Boloor}.The main challenge here is dealing with the over-provisioning so that does not compromise the profitability of the Cloud Provider. This profitability is named Business Objective Level (BLO) \cite{fish:Tao Q}.

\begin{figure}[!t]
\centering
\includegraphics[width=2.5in]{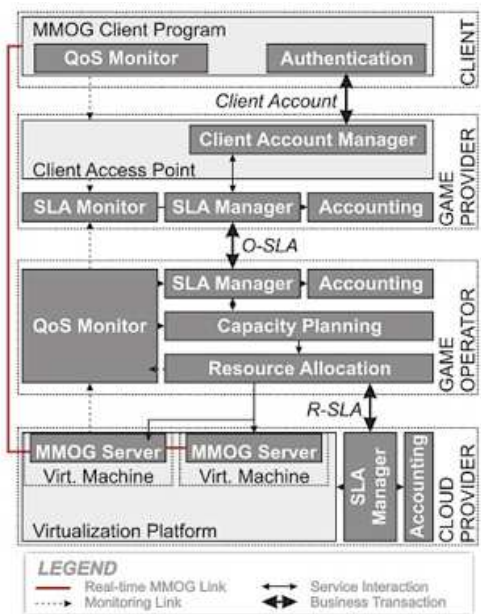}
\caption{Virtual SLA among system elements \cite{fish:Nae}.}
\label{fig_slaamoncomps}
\end{figure}

According to Nae et al. \cite{fish:Nae}, the risk of not achieving the BLOs refers to the absence of keeping QoS under control. Thus, the author suggests a model that creates a virtual SLA, delegating for the system elements the task of ensure QoS. That model is shown on Figure \ref{fig_slaamoncomps}.

Also, in an attempt to ensure the QoS, Boloor \cite{fish:Boloor} proposes a model that uses Weighted Round Robin (WRR) and First In First Out (FIFO) algorithms to scheduling data that come from servers geographically disposed. That approach pretends to organize the data requests to be delivered respecting limits of SLA terms.

In this study were identified evidences about the practice of submitting false information (better response time, high availability) to call the attention of the Customer \cite{fish:Tao Q}. Furthermore, some authors argue that lack of clarity in terms of the contracts \cite{fish:Kandukuri}\cite{fish:Boloor}.

This way, there are some suggested questions to perform before to agree on a contract \cite{fish:Kandukuri}: 

\begin{itemize}
\item In terms of availability (99.9\%), the Cloud provider can sign an SLA?; 
\item What happens when a breach of contract?;
\item How do I get my data at the end of the contract, what kind of data will be returned?.
\end{itemize}

Regarding the data privacy, Kandukuri et al. \cite{fish:Kandukuri} suggests the questions as follow:
\begin{itemize}
\item What data security level is implemented at the Physical Layer and the Network Servers?;
\item What about investigation Support?;
\item How much safe is data from Natural disaster?
\item How much trusted is Encryption scheme of Service Provider?
\end{itemize}

\subsection{RQ3 - What are the Cloud Computing social impact?}

About Social Impacts, the evidences point that Cloud Computing presents conflicts against Laws \cite{fish:Taylor} \cite{fish:Doelitzscher} \cite{fish:Udo}.  In \cite{fish:Taylor}, the author affirms that due to nature of Cloud Computing (distributed and flexible), to collect evidences about facts is a hard work. This is because a data inserted on Cloud environments could be encrypted before of its entering in Cloud platform, for example. Or, the stored data could be globally disposed among countries. Then, even though an authority solicits information about a fact, it will be difficult to track the data.

On Germany, the Federal Data Protection \cite{fish:ChoBrian} in section 11 describes that Customer, in this context, must know the location of your data. But,  Cloud Computing can put a data on anywhere because its nature. But, that problem is not isolated.

In \cite{fish:Zhou}, other conflicting aspects between Laws and Cloud Computing are presented on Acts such as HIPPA (Health Insurance Portability and Accountability), ECPA (Electronic Communications Privacy), and UPA (USA Patriot). Because those facts, some authors \cite{fish:Doelitzscher} \cite{fish:Udo} suggest changes both country Laws and Cloud Computing.

For United Nations/American Society for Public Administration , e-Govern-
ment is usage of the Internet and the world-wide-web for delivering Government information and services to citizens \cite{fish:UNASPA}. 

\begin{table}[hbp]
\renewcommand{\arraystretch}{1.3}
\caption{Cloud advantages from interaction perspective for e-Gov \cite{fish:Zissis}.}
\label{table_eGovAdv}
\centering
\begin{tabular}{| p{3.0cm}| p{8.0cm}|  }
\hline
\bfseries Criteria & \bfseries Product\\
\hline
Efficiency &- Provides uniform access to data and applications

\\
\hline
Effectiveness &- Improves data quality
\newline - Improves quality of services
\\
\hline
Strategic benefits&- Provides uniformity of solution
\newline - Introduces new services
\newline - Integrates existing infrastructure deployments
\\
\hline 
Transparency &- Constant evaluation and control of services and application usage, reduction of expenses
\\
\hline
\end{tabular}
\end{table}

 Zissis and Lekkas \cite{fish:Zissis} affirm that the migration of public datasets to the Cloud Environment should produces positive results both for the government and for citizens. In addition to reducing the delays of procedures of public layer, the authors present the benefits described in Table \ref{table_eGovAdv}.

Besides the advantages presented by Zissis and Lekkas \cite{fish:Zissis}, the adoption of Cloud Computing by the Government can foster the creation of a new market geared to this context.

Another point about Cloud Impacts is that with the creation, the increasing use of technologies such as \textbf{Hadoop MapReduce}\footnote{http://hadoop.apache.org/} and \textbf{Virtualization}, the possibility of working with out-sourced APIs (Application Programming Interface) and computing resources. Cloud Computing brings a emergent necessity of new professional profiles \cite{fish:Babu} \cite{fish:Hutchinson}.

An interesting viewpoint was found in respect to Green IT and Cloud. In Baliga et al. \cite{fish:Baliga}, the authors reported that energy saving donÕt occur for all cases. In that work, the authors relate that under some circumstances, Cloud Computing can consume more energy than conventional computing. 

The reason for this observation comes from the amplitude of the study \cite{fish:Baliga}. Therefore, in addition to take into account the datacenter environment, the authors also took into account the energy spent to maintain the entire infrastructure until the end customer. Thus, both the internal and the external environment were taken into account.

\subsection{What are the challenges found regarding to service conception on Cloud Computing environments?}

For Cadan et al\cite{fish:Cadan}, SaaS architectures can be classified into four maturity levels in terms of their deployment customization, configurability, multi-tenant efficiency, and scalability attributes.

At first level of maturity, the customer has its own application instance hosted on Cloud server. Customer migration from his traditional non-networked or client-server application to this level of SaaS typically requires the least development effort.

In second level, the keyword is \textbf{metadata}. With a configurable metadata is possible provides a flexible application instance for each customer. This approach allows the vendor to meet the different needs of each customer through detailed configuration options, while simplifying maintenance of the common code base \cite{fish:Cadan}.

The third level refers to multi-tenant efficiency. In that point all customers stay on unique instance of the application, but each one is treated such as \textbf{tenant}. This approach enables potentially more efficient use of server resources without any apparent difference to the end user.

 For  \cite{fish:Cadan}, in fourth and last level explicit scalability features are added through a multi-tier architecture supporting a load-balanced\footnote{http://www.f5.com/glossary/load-balancing.html} farm of identical application instances, running on a variable number of servers.

According to  Bonetta and Pautasso \cite{fish:Bonetta} the dynamic nature of Cloud Computing and its virtualized infrastructure pose new challenges in term of application design, deployment, and dynamic reconfiguration. In that work, the authors presented a novel parallel programming model named \textbf{Liquid Architecture}.

The approach pretends to builds application services that can be transparently deployed on multicore and Cloud execution environments. Behind of concept, the approach uses fundaments of REST (Representation State Transfer) and loosely coupled components to exchange messages and creates an auto adaptive environment.

\begin{figure}[!t]
\centering
\includegraphics[width=2in]{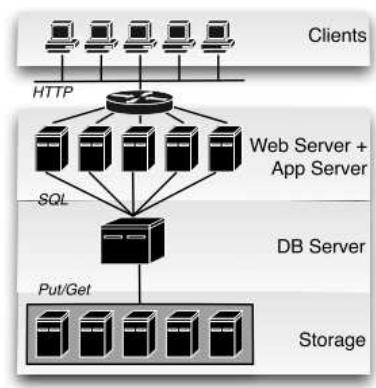}
\caption{Classic Database Architecture \cite{fish:Kossmann} }
\label{fig_classicArch}
\end{figure}

In  \cite{fish:Kossmann} are presented examples of infrastructure of Cloud environments. The Figure~\ref{fig_classicArch} shows the classic database architecture.

 For Classic Architecture, as just one server composes DB Server layer it can be a bottleneck in future. To solve problem, companies acquire expensive servers to replace him. To reach scalability in that model, companies can insert new Solids Disks on Storage Layer. 

\begin{figure}[hbp]
\centering
\includegraphics[width=2in]{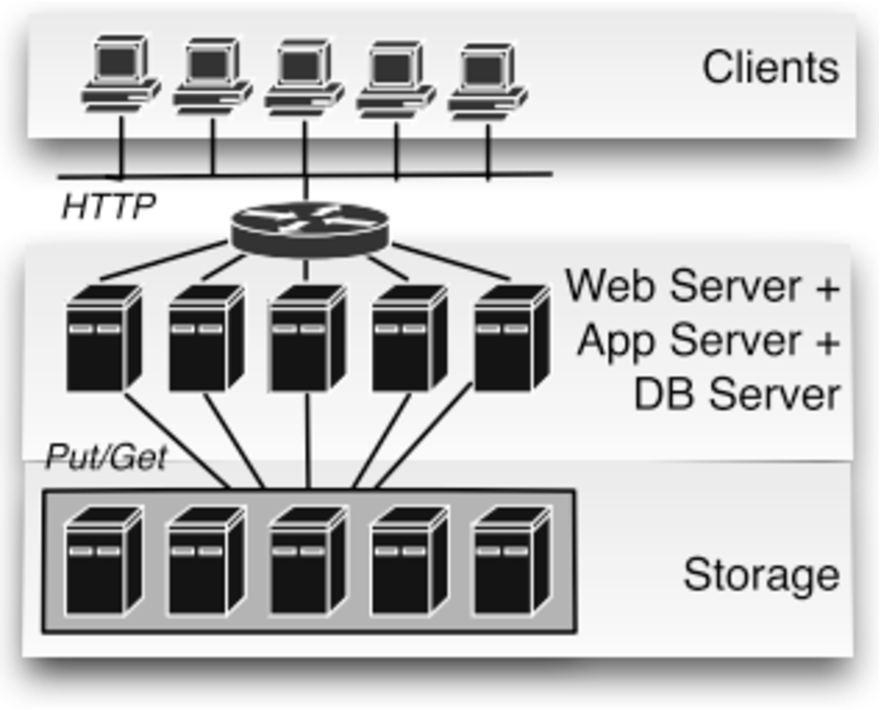}
\caption{Distributed Control \cite{fish:Kossmann} }
\label{fig_DistributedControl}
\end{figure}

Still, Kossmann \cite{fish:Kossmann} presents others variations of that architecture (see Figure~\ref{fig_classicArch}). However, the variation more suitable to Cloud is presented on Figure~\ref{fig_DistributedControl}, Distributed Control. In that model, the Cloud has high scalability. However, that aspect has a cost.

With physical and logical growing on system, the Cloud environment cannot insuring Resilience, Consistency and Availability. For that, consistency is adequate to an enough level. In technical terms the impact occur on ISOLATION property of ACID \cite{fish:Chapple}.

Also, it is important mention that a common referred work to guide stakeholders in the development of architectures in the Cloud was the RESERVOIR framework \cite{fish:Chapman}\cite{fish:LinShih} \cite{fish:Elmroth} \cite{fish:Costa}. That composed by a multi-tier model and its key difference is the ability to treat federate environment across different sites. 
\begin{figure}[hbp]
\centering
\includegraphics[width=4.5in]{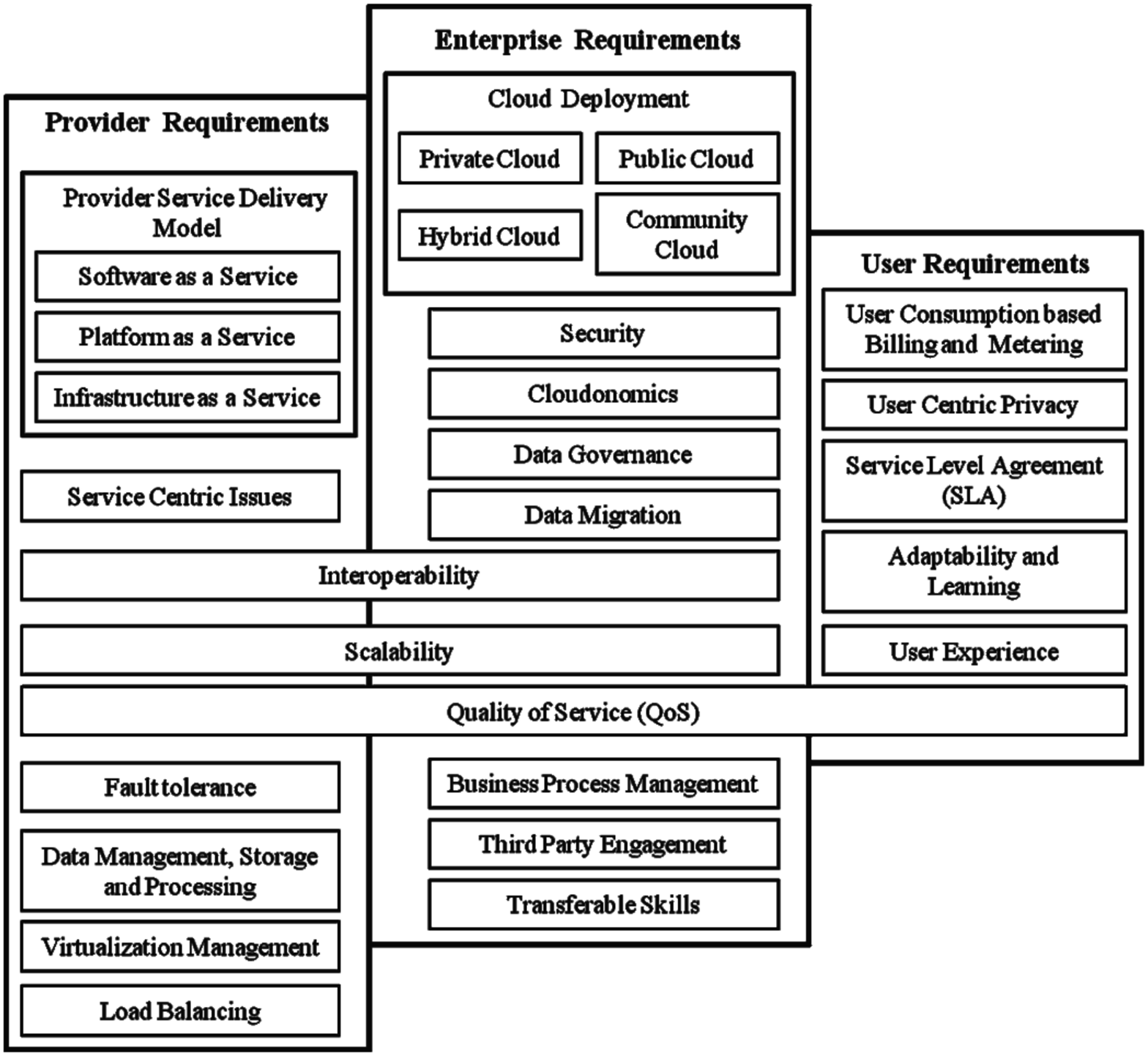}
\caption{Requirements for development of  Cloud Computing environments \cite{fish:Rimal}}
\label{fig_requirements}
\end{figure}

For Rimal et al. \cite{fish:Rimal}, there are requirements that must be taken into consideration with regard to the development of a Cloud Computing environment. In that work, the authors divide the requirements in the views of the Cloud Provider, organizations and end-user. As shown on Figure \ref{fig_requirements}.

In the context of organizations (Figure \ref{fig_requirements}), when the authors refer to \textbf{Cloudonomics}, it is dealing with rules that must be considered in order to provide Cloud Computing benefits over conventional datacenters\footnote{http://www.businessweek.com/technology/content/sep2008/tc2008095\_942690.htm}.

\subsection{RQ5 - What are the main challenges regarding to the Elastic property?}

\begin{figure}[hbp]
\centering
\includegraphics[width=3.6in]{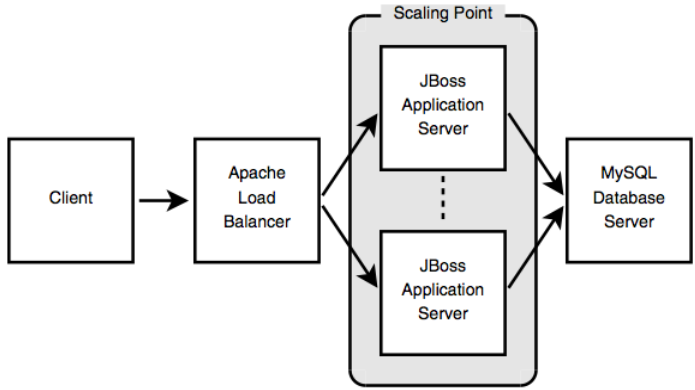}
\caption{Load Balancer and Scalling Point \cite{fish:Dutreilh}}
\label{fig_loadBalancing}
\end{figure}

Among evidences found in this work,  the main used techniques to treat Resource Allocation problems is the Load Balancing \cite{fish:Mehta} \cite{fish:Dutreilh} . The most of authors use a load balancer mechanism to manage instances of application servers, as follow on Figure \ref{fig_loadBalancing}.

Another solution to Resource Allocation is proposed by \cite{fish:LiQiang}. That author uses FeedBack mechanism for deliver computing resource. The mechanism is based on CPU, I/O and RAM memory usage. 

\begin{figure}[hbp]
\centering
\includegraphics[width=3.7in]{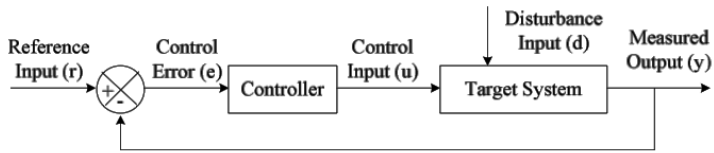}
\caption{A standard feedback control system \cite{fish:LiQiang}}
\label{fig_feedbackSystem}
\end{figure}

A standard model of that mechanism is shown in Figure~\ref{fig_feedbackSystem}.

The \textbf{Target System} (see Figure \ref{fig_feedbackSystem}) is the computing system (in Cloud environment) managed by the \textbf{Controller}. In order to achieve a desired objective of the Target System (accurate provisioning), the Controller dynamically regulates the environment based on feedbacks based on difference between \textbf{Reference Input} and \textbf{Measured Output}.

For Nair et al.\cite{fish:Nair}, a \textbf{broker system} model can be a good way to deliver service according to SLA terms. It is because a component on Cloud environment is responsible to allocate resources respecting contractual terms. So, when a customer throws a request to Cloud provider, a broker system is responsible for allocate computing resource from local farm. Otherwise, its resource will be allocated from a third company such as Amazon AWS. Thus, causing \textbf{cloud bursting}.

Another proposal to assist in resource allocation is the \textbf{Prediction} \cite{fish:Caron}. The prediction consists in allocating resources, based on future needs of Cloud environment.

For Caron et al. \cite{fish:Caron}, besides Prediction, a Cloud needs reduce its time to allocate resource. So, that work proposes Prediction based on similarity of occurred historic. In other words, a mechanism needs analyze scene already occurred on Cloud and prepare it to execute allocation on future opportune time.

\subsection{RQ6 - What are the problems and solutions about data storage?}
Because concentrating of high quantity of requests to storage system, \textbf{I/O} is an evident problem on Cloud Computing \cite{fish:Wang} \cite{fish:Sivathanu}. Some authors \cite{fish:Ke Xu} \cite{fish:Cheng} proposes P2P model as alternative to solve that problem.

For \cite{fish:Yusuke}, the solution for I/O problem is an approach based on reservation time to requests.  When the user accesses the storage system during the reserved time, the requested performance is guaranteed because the storage system allocates the resources according to the reservation, and prioritizes I/O requests for the reserved access.

Also, alternatives on \textbf{data compression} have been proposed, not only the problem of I/O, but the use of the bandwidth \cite{fish:Nicolae}.

Another concern is the \textbf{Big Data}\footnote{http://www-01.ibm.com/software/data/bigdata/}. With a lot of data generated by people, systems, and companies, Cloud Computing comes such as solution to storage that data. However,  Kozuch et al. \cite{fish:Kozuch} proposes that Cloud storage systems utilize location-aware mechanism to store data. That study refer to project TASHI\footnote{http://incubator.apache.org/tashi}, supported by Apache incubator.

Hadoop MapReduce is framework for writing applications that rapidly process vast amounts of data in parallel on large clusters of compute nodes. That framework has been used in other studies found in this work \cite{fish:Wang} \cite{fish:Wang Yuxiang}. 

However, Kozuch et al. \cite{fish:Kozuch} alerts that Hadoop MapReduce framework can impacts on software development, case some stakeholder wish migrates its software to that framework. 

For guarantee the integrity of data on Cloud, the environment needs to adopt a Proof Of Integrity (POI) protocol \cite{fish:Kumar}\cite{fish:Zheng}. Such protocol prevent the Cloud storage archives from misrepresenting or modifying the data stored without the consent of the data owner by using frequent checks on the storage archives. However, Kumar and Sexena\cite{fish:Kumar} emphasize that POI protocol should be used with caution because of the possibility of overhead on the system.

Also were found efforts to insert On-Line Analytical Processing (OLAP) systems on Cloud environments \cite{fish:Wang Yuxiang}. In that case the authors extend the MapReduce framework \cite{fish:Hadoop} . Furthermore, Johnson \cite{fish:Johnson} presents suggestions about ways to perform SQL queries on Cloud.

\subsection{RQ7 - How is performed the resource usage monitoring on Cloud Computing?}
For Spring \cite{fish:Spring}, the monitoring is a big ally to SLA (Section \ref{sla}) and Security (Section \ref{sec}) on Cloud environments.  In that work, the author presents what must be the amplitude of monitoring control for Cloud Providers, as follow on Table \ref{table_monitoringBound}.

\begin{table}[hbp]
\renewcommand{\arraystretch}{1.3}
\caption{Cloud Providers Monitoring control  \cite{fish:Spring}}
\label{table_monitoringBound}
\centering
\begin{tabular}{| p{2.0cm}| p{1.5cm}| p{1.5cm}| p{1.5cm}| }
\hline
\bfseries  Level & \bfseries SaaS & \bfseries PaaS & \bfseries IaaS\\
\hline
Facility(Fisical) & X &  X & X\\
\hline
Network & X &  X & X\\
\hline
Hardware & X &  X & X\\
\hline
O.S. & X &  X & ?\\
\hline
Application & X &  - & -\\
\hline
User &  - &  - & -\\
\hline
\end{tabular}
\end{table}

In the Table \ref{table_monitoringBound}, (\textbf{X}) refer to control performed through monitoring,(\textbf{--}) represents unreachable elements, and (\textbf{?}) means the control of monitoring scheme depends on type of implementation on element.

For Elmroth and Larsson \cite{fish:Elmroth}, two approaches can be used in monitoring context. In the first, the monitoring system is charged for observes behaviors on infrastructure based on Hard disk, RAM memory and Virtual Machines resource usage. The second, is charged for observes point in applications modules such as quantity of users logged or life time of threads.

This Research Question also intents to acquire tools about monitoring. Thus, the tools referred by authors are presented on Table \ref{table_monitoringTools1}.

\begin{table}[hbp]
\renewcommand{\arraystretch}{1.3}
\caption{Used tools and frameworks for monitoring on Cloud Computing.}
\label{table_monitoringTools1}
\centering
\begin{tabular}{| p{2.6cm}| p{8.5cm}|}
\hline
\bfseries  Frame./Tool & \bfseries Description \\
\hline
GrenchMark& Framework for performance testing and analysis, system functionality testing, and comparing setting. Initially, the project was used for grid computing, but evidences its use in Cloud Computing. \newline
\textbf{Project site: grenchmark.st.ewi.tudelft.nl}.
\\
\hline
C-METER &  An extension of the project GrenchMark adapted to Cloud Computing \cite{fish:Yigitbasi}. \\
\hline
Monalytics &  Framework for monitoring data, taking into account the scalability of the Cloud. The author  divides the Cloud environment into zones, allowing the monitoring only on interesting areas\cite{fish:Kutare}\\
\hline
\end{tabular}
\end{table}

\begin{table}[!t]
\renewcommand{\arraystretch}{1.3}
\centering
\begin{tabular}{| p{2.6cm}| p{8.5cm}|}

\hline
CloudClimate &Monitoring of Cloud Computing online. Agents are installed in environments Cloud (eg Amazon) and perform tests of performance, sending the report to the project site. 
\newline
\textbf{Project site:www.cloudclimate.com}.
\\
\hline
Grid Monitoring Architecture (GMA) & This framework is commonly cited as base to others monitoring extensions. \cite{fish:Elmroth}\cite{fish:Huang}\\
\hline
Push\&Pull model (P\&P) & A approach based monitoring actions of the type Pushing and Pulling.\cite{fish:Huang}.\\
\hline
File System in User Space (FUSE)& used to monitor I/O to the file system.
\newline
\textbf{Project site:fuse.sourceforge.net}.
\\
\hline
Joulemeter& Monitoring of energy consumption of the Virtual Machines. \cite{fish:Kansal}\\
\hline

\end{tabular}
\end{table}

\subsection{RQ8 - Which are the main security challenges?}\label{sec}
According Wayne A. Jansen (NIST)\cite{fish:WayneA}, the main challenges regarding security and privacy can be divided into some sub-categories. Thus, the others evidences were divided based on those categories.
\subsubsection{Trust}
It is important that both the Customer and Cloud Provider understand that by adopting the paradigm of Cloud Computing, the organization (Customer) delegates control of security system to the service provider.

Thus, to avoid creating gaps in environment, security policies, monitoring, processes and control techniques must be applied on Cloud Provider \cite{fish:Zhang Xuan}\cite{fish:Mounzer}.

\subsubsection{Architecture}
Security architecture challenges, are linked directly with the care of the elements that compose it.

A Cloud environment is composed of software components and Hardware. Virtual machines (VMs) typically serve as the abstract unit of deployment and are loosely coupled with the Cloud storage architecture. Moreover, the applications carried in Cloud are usually created by intercommunicating among components of the environment.

Thus, were found identified proposals of safety models for Cloud Computing \cite{fish:Jing Xue}\cite{fish:Dai}, efforts against \textbf{intrusion} \cite{fish:Hai Jin}\cite{fish:Lee Jun},  virtual networks security model \cite{fish:Wu Hanqian}, and  \textbf{patching}\footnote{software updating system to void vulnerabilities related deprecated versions} model \cite{fish:Zhou Wu}.

\subsubsection{Identity Management}
Data sensitivity and privacy of information have increasingly become a concern for organizations, and unauthorized access to information resources in the cloud is a major issue. One reason is lack  \textit{cloud-driven} frameworks for that \cite{fish:WayneA}.

\begin{figure}[!t]
\centering
\includegraphics[width=4in]{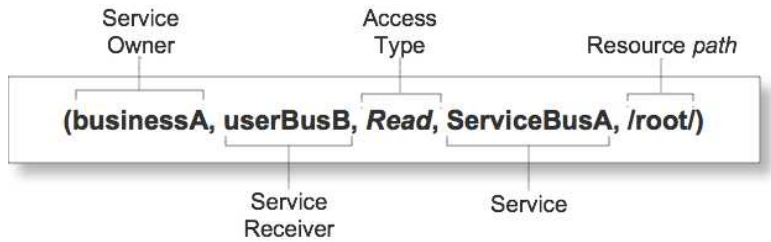}
\caption{Cloud Federation Access Control API  \cite{fish:Calero}}
\label{fig_tuple}
\end{figure}

In context, \cite{fish:Calero} proposes a Access Control API for \textbf{Cloud Federations}, where a tuple is adopted for each stakeholder on federation. As follow on Figure \ref{fig_tuple}.

Carelo et al. \cite{fish:Calero} adopts a RESTfull approach where each resource on federated Cloud is accessed according to a 5-tuple structure and a hierarchy. For example, an 5-tuple like (Paul, Mariah, Read, CloudStorage, Ò/root/Ó) means that Paul informs to system that Mariah has access for read the folder Ò/root/Ó of CloudStorage service. That API (application Program Interface) must be a common language for whole environment.

\subsubsection{Software Isolation}
Multi-tenancy in Cloud Computing is typically done by multiplexing the execution of VMs from potentially different users on the same physical server \cite{fish:WayneA}.

Thus, if a attack occur over a user, the Cloud Provider must reach a security level, which isolate the problem just for that client. So, other users of same server can perform yours transactions without interference. 

For then, understanding the use of \textbf{virtualization} by a Cloud Provider is a prerequisite to understanding the risks involved \cite{fish:WayneA}. 

\subsubsection{Data Protection}
Data stored in the cloud typically resides in a shared environment collocated with data from other customers. Organizations moving sensitive and regulated data into the Cloud, therefore, must account for the means by which access to the data is controlled and the data is kept secure \cite{fish:WayneA}.

\begin{figure}[!t]
\centering
\includegraphics[width=3.3in]{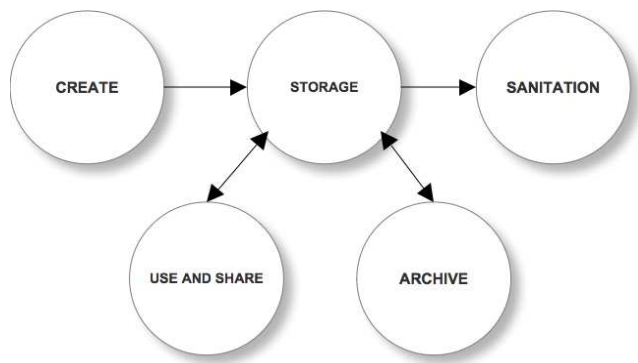}
\caption{Modelo de ciclo de vida de dados para Cloud Computing \cite{fish:Yu Xiaojun}}
\label{fig_lifecicle}
\end{figure}

Thus, Yu and Wen \cite{fish:Yu Xiaojun} propose a data life cycle model in order to follows all stages of user data, as show on Figure \ref{fig_lifecicle}.

The model (Figure \ref{fig_lifecicle}) intents extend security solutions to other stages of data, besides of storage. Also, other effort to protect data related to data mining and cryptography \cite{fish:Sigh Meena}, RSA algorithm usage \cite{fish:Jianhong}, and  kNN \textit{queries} with cryptography support \cite{fish:Wong} were found.

\section{Analysis of the Results and Mapping of Studies}

This section represents the analysis that we performed about the 301 studies found in this research. That effort enables us to present the number of 
\begin{figure}[hbp]
\begin{minipage}[b]{0.5\linewidth}
\centering
\includegraphics[width=2.3in]{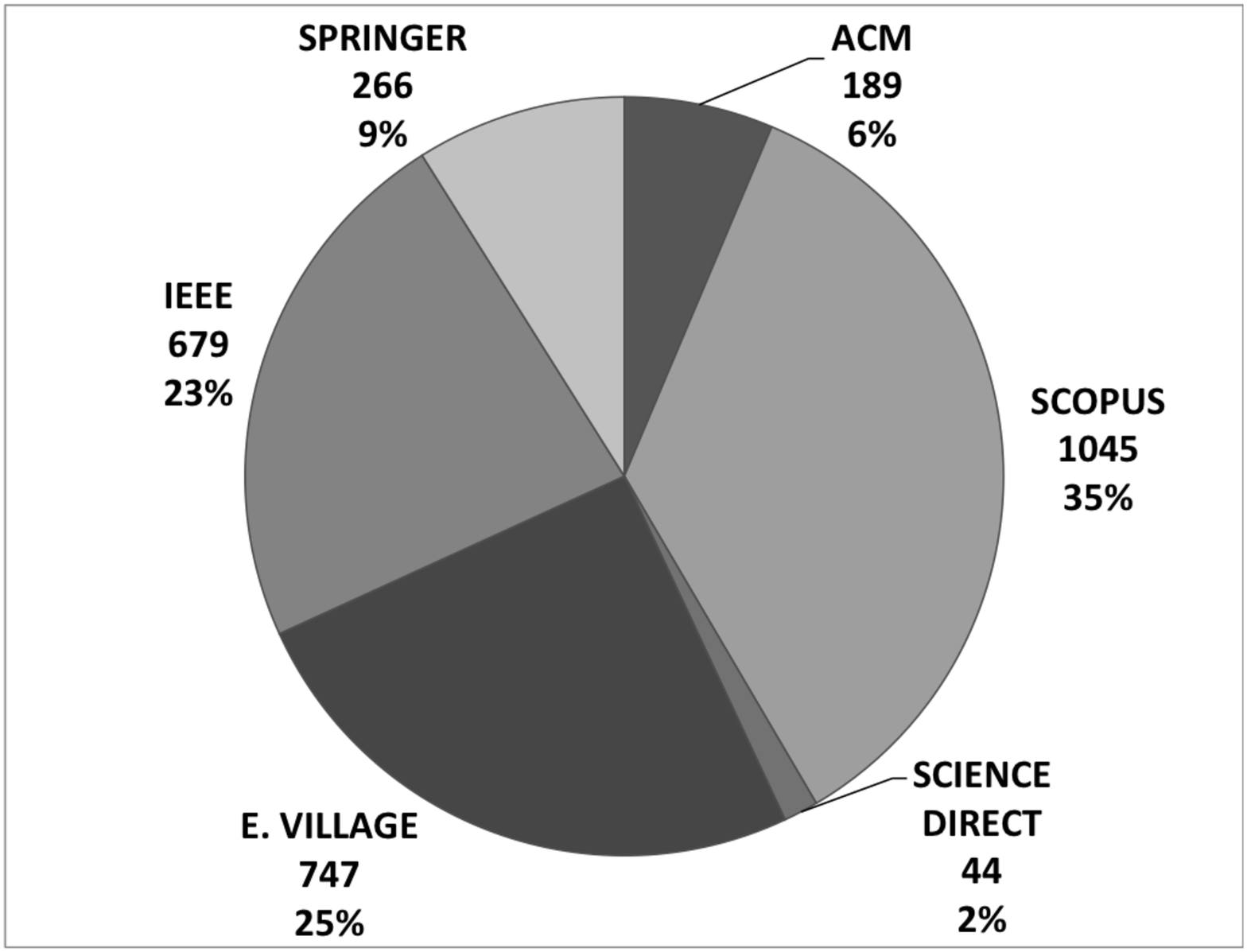}
\caption{Search Engines results before filters.}
\label{fig:results1}
\end{minipage}
\hspace{0cm}
\begin{minipage}[b]{0.5\linewidth}
\centering
\includegraphics[width=2.3in]{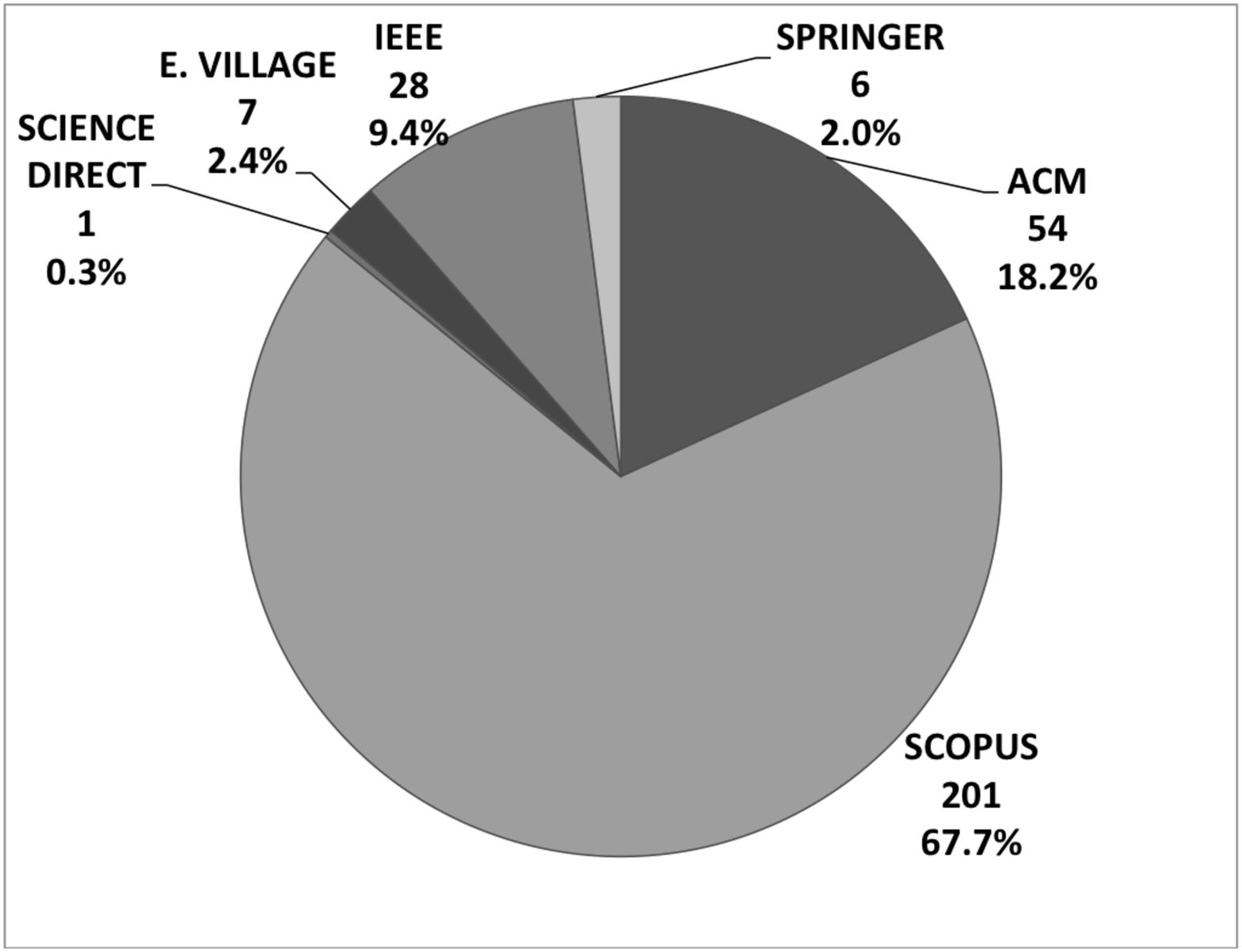}

\caption{Search Engines results after filters}
\label{fig:results2}
\end{minipage}
\end{figure}studies tabulated in each category defined in this work.

Thus, it is possible to identify what have been emphasized in past research and thus to determine gaps and opportunities for future research \cite{fish:KPetersen}.

Among search engines, the Scopus presented major efficiency for scope of this work(see Figure \ref{fig:results2}). That information, may lead to some interesting series of studies to evaluate the efficiency of the automated search engines.

\begin{figure}[!t]
\centering
\includegraphics[width=3.6in]{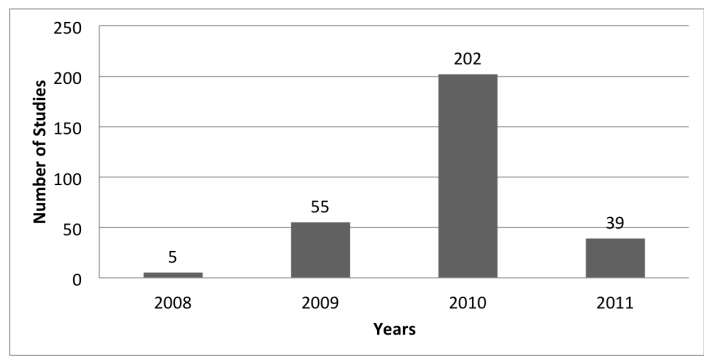}
\caption{Selected studies by years.}
\label{fig_papersByYear}
\end{figure}

The search through automated engines did not impose a time constraint regarding the publishing year. Then, for this research the selected studies were published from 2008 to 2011. 

Because the needed time to index studies, the Digital Libraries presented a minor quantity of works on 2011. However, the manual research was performed during remain year in order to obtain more evidences about addressed issues and mitigate other suggestions from the SMaRT research group members.

Also, due the emergence nature of Cloud Computing area, the choice of conferences and letters was conduced by quality of involved institutes (ACM, IEEE, Springer, and Elsevier) and experience of researchers on SMaRT research group.

\begin{figure}[hbp]
\centering
\includegraphics[width=2.7in, height=2in]{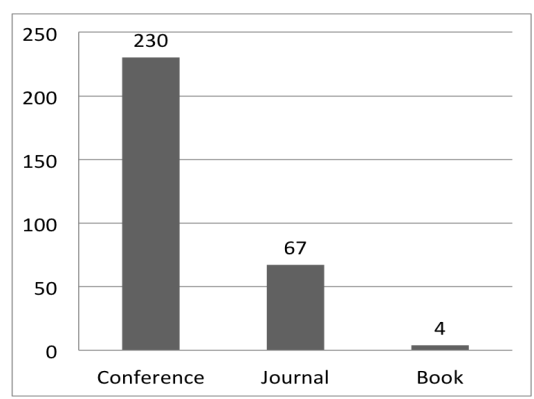}
\caption{Amount of studies by source}
\label{fig_confsvsjournal}
\end{figure}

In this research, the conferences presented the most of selected studies (see Figure \ref{fig_confsvsjournal}). Among sources, four letters are Springer Lecture Notes in Computer Science. Also, a list with 10 first sources of conferences and journals are available on Appendix A.

A interesting point identified was the recent edition of main targets: IEEE International Conference and Workshops on Cloud Computing Technology and Science, \textbf{4th edition} and International Conference on Cloud Computing, \textbf{5th edition}.  This way, characterizing a recent interest in the academy.

\begin{figure}[!t]
\centering
\includegraphics[width=3.3in]{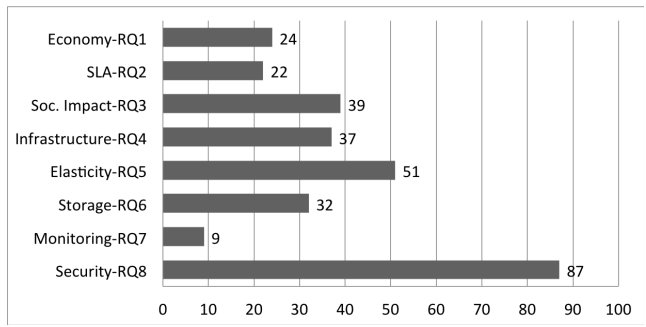}
\caption{Number of studies by Research Question}
\label{fig_papersByRQ}
\end{figure}

In Figure \ref{fig_papersByRQ} are presented the distribution of studies by Research Question. The RQ7 (Monitoring) had the lowest number of studies due to the scope of the question. 

\begin{figure}[hbp]
\centering
\includegraphics[width=3.6in]{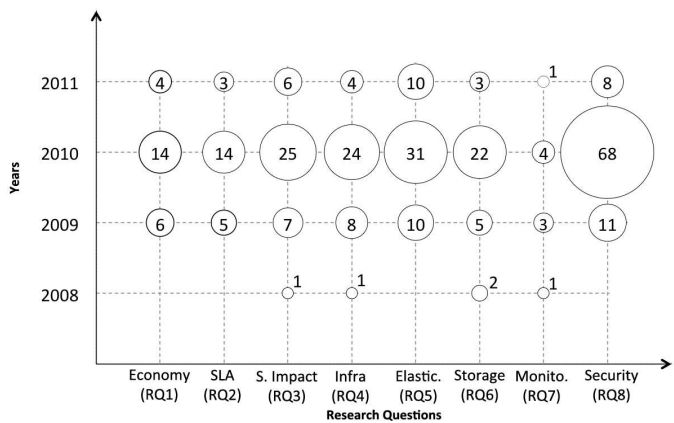}
\caption{Classification by year of studies in Research Questions}
\label{fig_rqByYear}
\end{figure}
On the other hand, the RQ8 (Security) had the highest number due to  activity of the authors about one of the most polemic issues on Cloud Computing. Also, the classification with references can be found on Appendix B.

By analyzing the studies of 2008 year, was possible identify the close relation of Cloud Computing concept with storage service. Thereby, Amazon AWS was the most referred Big Player in this stage of the approach. Over the years new players have been referenced, including Salesforce, Rackspace, and Google.

Also, it is important to emphasize that in this year was possible to identify the interest of the healthcare sector on the proposed cost reduction of Cloud Computing versus Grid Computing solutions of period. In this case, solutions for detection of infectious disease outbreaks as \textbf{University of Pittsburgh RODS Laboratory (Real- time Outbreak and Disease Surveillance)}.

\begin{figure}[!t]
\centering
\includegraphics[width=3in]{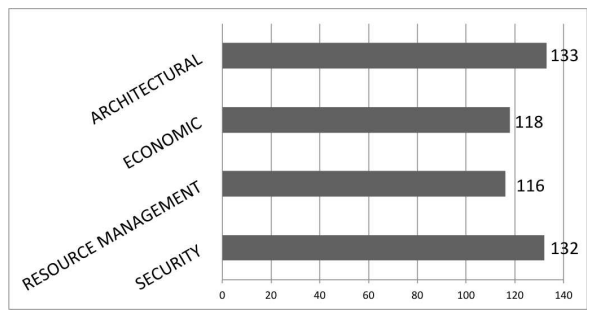}
\caption{Number of studies by Aspects}
\label{fig_papersByAS}
\end{figure} 

Still, beyond the classification of Research Questions, the studies were classified according to  aspects addressed in its content (see Figure~\ref{fig_papersByAS}).

\begin{figure}[hbp]
\centering
\includegraphics[width=3.6in]{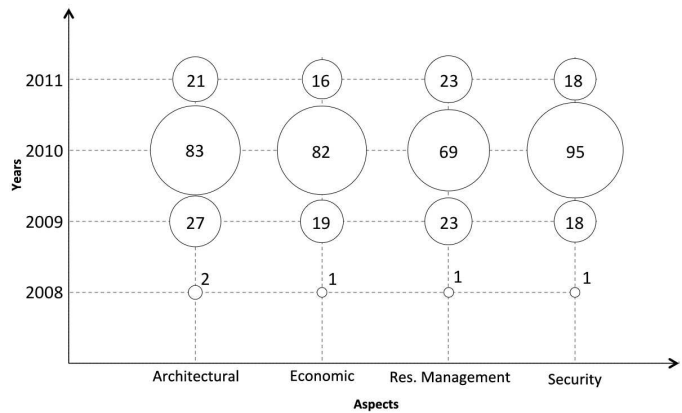}
\caption{Aspect by Year}
\label{fig_ASbyYear}
\end{figure}
 \hvFloat[
 floatPos=!t,
 capWidth=h,
 capPos=r,
 capAngle=90,
 objectAngle=90,
 capVPos=c,
 objectPos=c]{figure}{\includegraphics[width=7.2in]{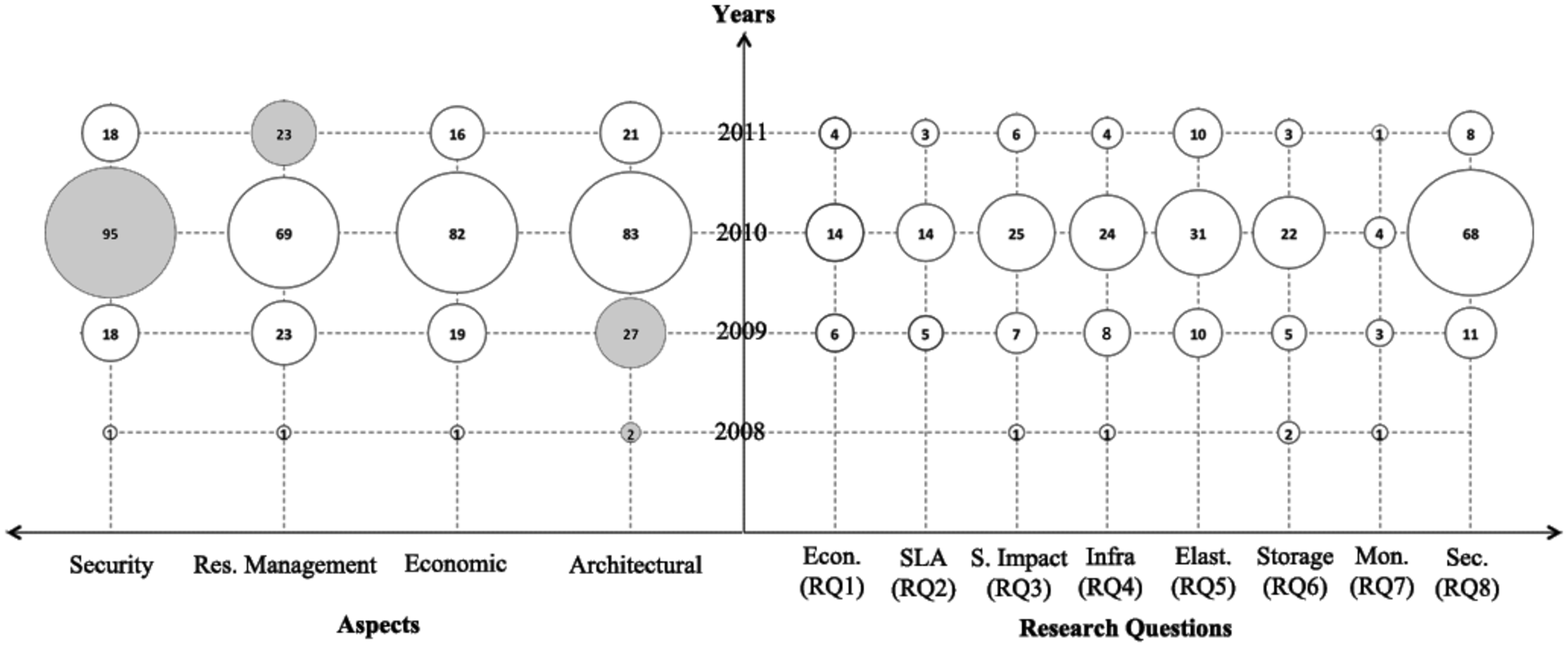}}%
{Major aspect influence by Year}{fig_XXXXAS}

 The aspects analyzed were: \textit{Architectural, Resource Management, Security, and Economic}, already described on Section \ref{method}. It is important mention a study can be classified on more than one aspect.

As shown on Figure \ref{fig_papersByAS}, all aspects are quite referenced among authors. Thus, demonstrating their influence in the design of studies. However, \textbf{Resource Management} aspect shown the lowest number of occurrence.

This is because the influence of Cloud Flexibility (Res. Management) was smothered by a security ``fever'' of 2010, depicted on Figure \ref{fig_XXXXAS}. On the other hand, the authors were quite influenced by the others aspects due to context of that year. 

However, the 2011 year presents a different scenario for Resource Management aspect. But, we waited for more evidences about performance, algorithms and strategies for Flexibility context of Cloud Computing.

We understand that one of causes for lack of more evidences about Flexibility on Cloud Computing is related to proprietary standards and business values of strategy. Because when the elasticity of the environment enables high quality of user experience and promotes the Cloud Provider's profit, it will be its difference in the market.

\begin{equation}
\label{formulafish}
R(RQ,x) = \frac {Aspect_{x}(RQ)*100} {\displaystyle{\sum_i Aspect_{i}(RQ)}}, 1\le i  \le 4
\end{equation}

In search for detailed composition of contents, we merge the Research Question and Aspects facet. Thus, for better understanding, the numbers were reported in percentages. For this, a simple calculation of percentage was used, as shown on expression \ref{formulafish}.

\begin{figure*}[hpb]
\centering
\includegraphics[width=4.6in,height=2.6in]{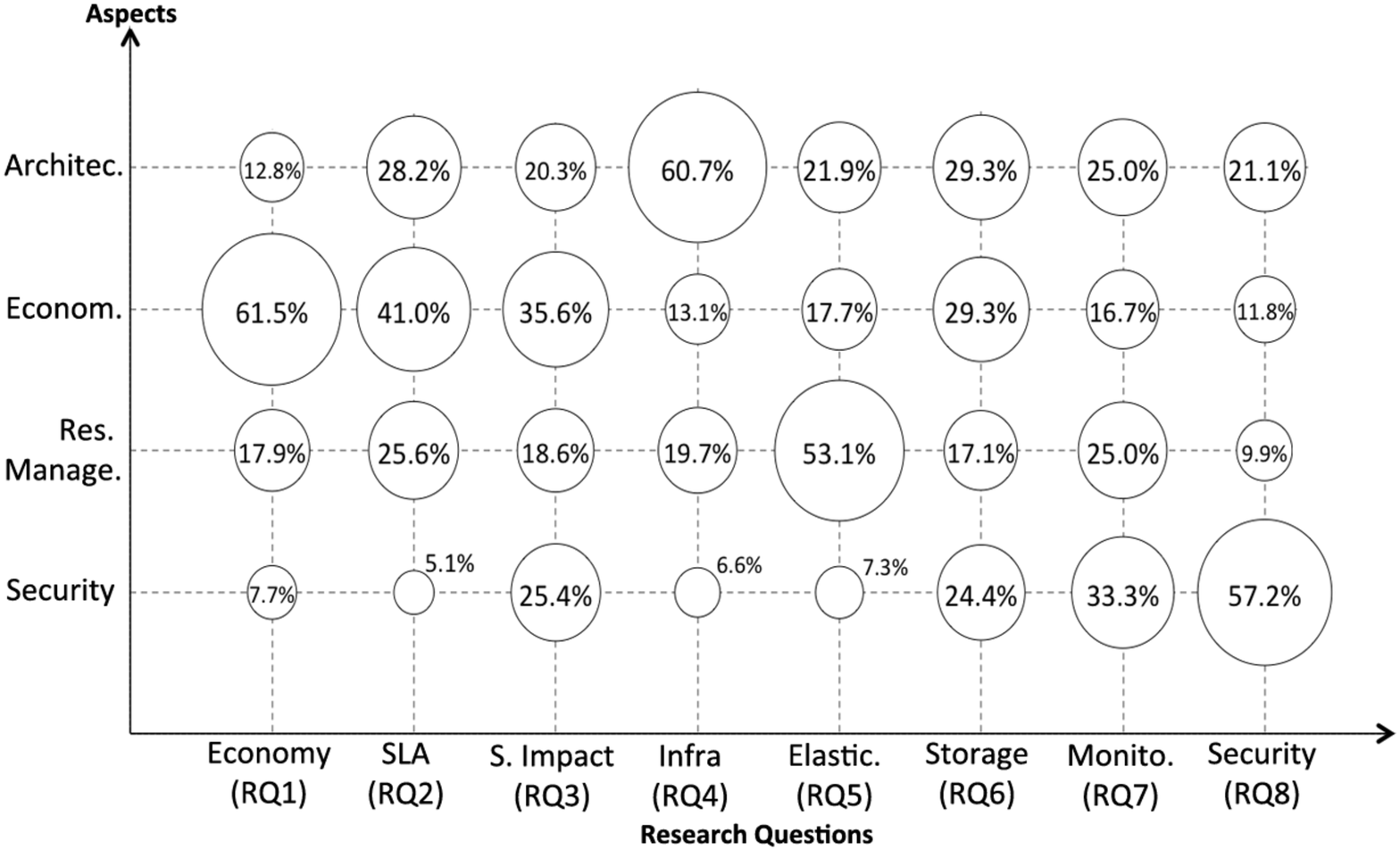}
\caption{Aspects Facet vs RQs Facet.}
\label{fig_percentCrossing}
\end{figure*}

That way, when merging the two Facets we found the composition of content for each RQ, as shown on Figure \ref{fig_percentCrossing}.

By analyzing the RQ8 studies, it can be stated that many efforts are being directed to security in order to mitigate its risks. Thus, making Cloud Computing an increasingly attractive option.

In RQ2 (SLA),  studies have emphasized the economic aspect, due to possible penalties  that a Cloud Providers can pay in consequence of not complying with the proposed terms of contracts, such as Level Quality of Service (QoS) \cite{fish:Boloor}. Thus, management of resources and the architectural aspect of Cloud Computing, was taken into consideration because of its close relationship with the problem.

Due to the attractive appeal of low cost of Cloud and new employment opportunities, the economic aspect was the predominant factor in RQ3 (soc. Impact). The aspect of security relates to privacy of data stored in the cloud and the centralization of critical data in a single place.

Also worth noting is the influence of security RQ7 (Monitoring). Even with a scope focused on the design of monitoring, the security factor was more evident.

\begin{figure}[hbp]
\centering
\includegraphics[width=3.6in,height=2.1in]{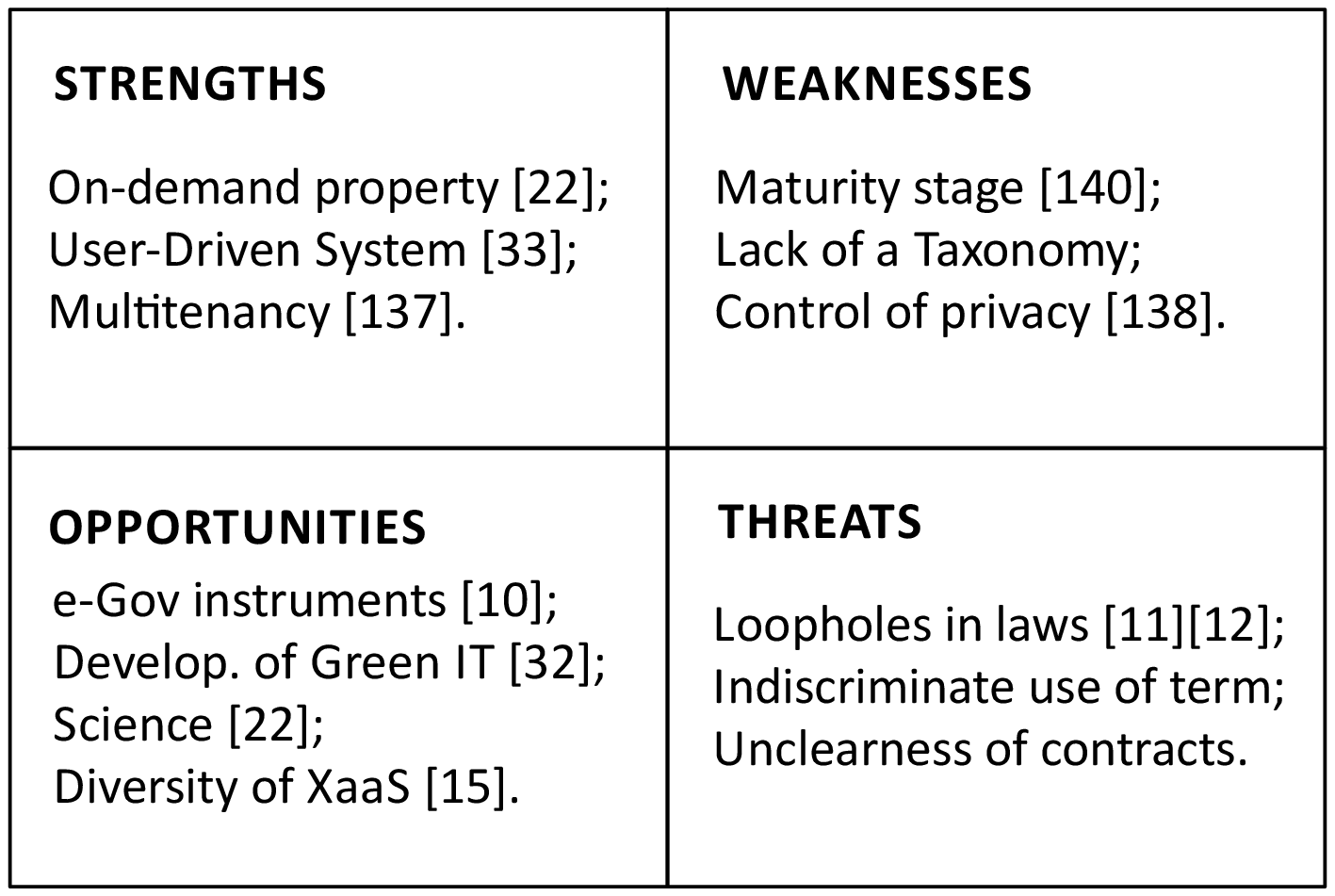}
\caption{S.W.O.T. analysis of Cloud.}
\label{fig_swot}
\end{figure}

A S.W.O.T. analysis was created about points that we understand to be important in this work (see Figure~\ref{fig_swot}). Among properties, During the development of keywording step, described in section \ref{method}, were identified the lack of a well-defined taxonomy. Therefore, our classification also extended to the contents of studies.

Multitenancy was inserted on strengths because your Longtail effect\footnote{http://www.wired.com/wired/archive/12.10/tail.html} for Cloud. In other words, a same infrastructure can meets with requirements both organizations and final users, besides to enables the Cloud Provider profit. For instance, in the same way of Google (gmail) and Salesforce (CRM)\footnote{http://www.salesforce.com}.

\begin{table}[hbp]
\renewcommand{\arraystretch}{1.3}
\caption{XaaS Found.}
\label{table_XAAS}
\centering
\begin{tabular}{| p{5cm} | p{6cm}|}
\hline
\bfseries Service & \bfseries Description\\
\hline
NaaS (Network as a Service) & Delivery bandwidth according with factors as QoS, Reliability and etc.\cite{fish:Fabio Baroncelli}.\\
\hline

\end{tabular}
\end{table}

\begin{table}[!t]
\renewcommand{\arraystretch}{1.3}
\centering
\begin{tabular}{| p{5cm} | p{6cm}|}

\hline
ASaaS (Application Software as a Service) & Software for High Performance scientific Computing \cite{fish:Hou_Zhengxiong}.\\
\hline
SCaaS (Supply Chain as a Service) & Software for supply chain on Cloud \cite{fish:Leukel}.\\
\hline

PasS (privacy as a service) & A outsourced service to encrypt data of clients \cite{fish:Itani}.\\
\hline
DaaS (Database as a Service) & Database on Cloud \cite{fish:Alzain}.\\
\hline
"SaaS BI" or BIaaS (Business Intelligence as a Service) & An conceptual framework of B.I. on Cloud \cite{fish:Tang}.\\
\hline
CaaS (Continuous Analytics as a Services) & Store analytical data on Cloud \cite{fish:Begnum}.\\
\hline
Travel Reservation as a Service & Reservation system with SOA architecture on Cloud \cite{fish:Namjoshi}.\\
\hline
Process as a Service & Governance of Process Runtime on Cloud \cite{fish:Wang MingXue}.\\
\hline
\end{tabular}
\end{table}

The possibility of provide new service models, Everything as a Service (XaaS), also becomes Cloud Computing attractive to stakeholders. We create a table with some services (beyond IaaS, PaaS, and SaaS) identified in this mapping (Table~\ref{table_XAAS}).

\begin{figure}[hbp]
\centering
\includegraphics[width=4.6in]{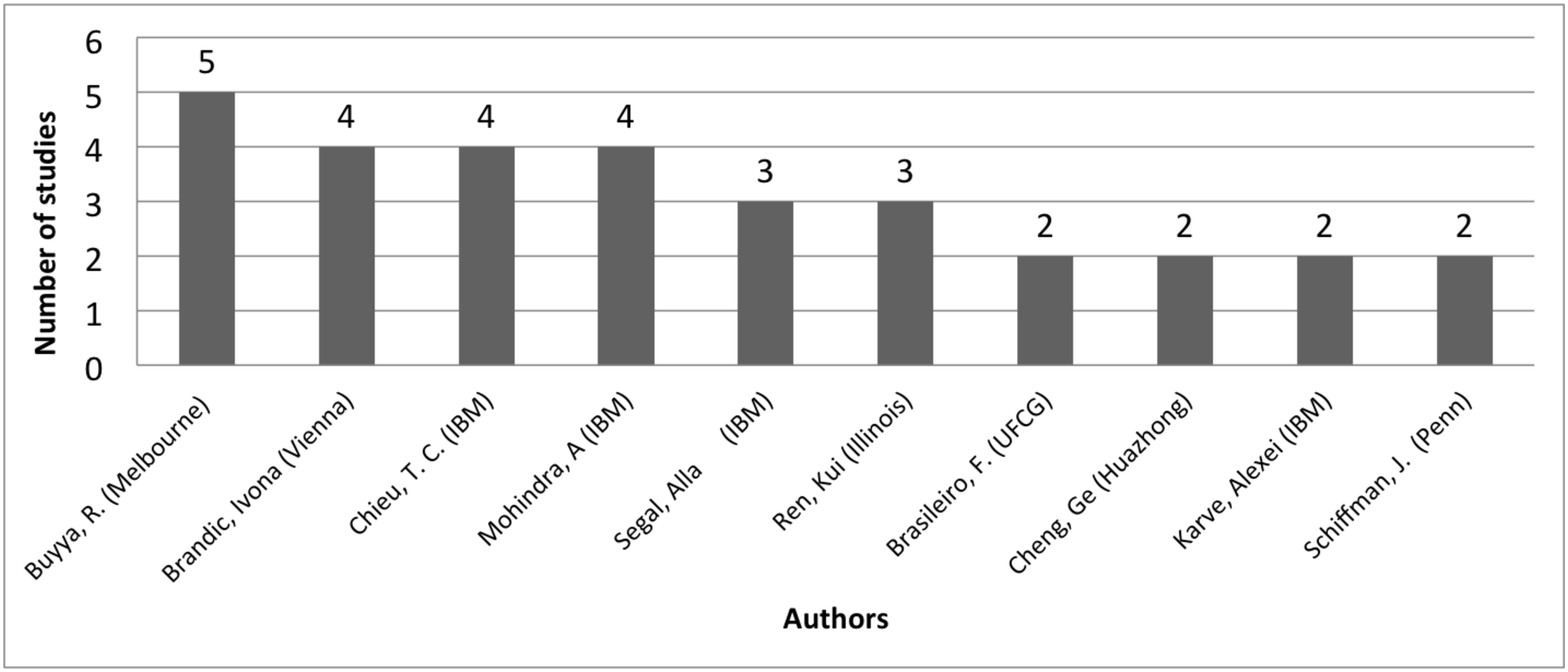}
\caption{Largest contributors.}
\label{fig_authors}
\end{figure}

This research registered 1246 authors. We identify the 10 largest contributors on Cloud context based on number of published studies.  As shown on Figure \ref{fig_authors}

 The highlights of this point was Rajkumar Buyya at University of Melbourne, Australia and for  IBM T.J. Watson Research Center, USA.

 Including, the Professor Rajkumar was a guest editor to first edition of 2012 of  IEEE Transactions on Parallel and Distributed Systems\footnote{Special Issue IEEE: http://goo.gl/4JiVM}.

 Also, were identified the focus of these 10 authors related to Research Questions. As shown on following Figure \ref{fig_authorsbyrq}, the most focused authors were Ren kui and Ivona Bradic. An interesting point is the participation of the IBM through Trieu Chieu.
\begin{figure}[!t]
\centering
\includegraphics[width=4.5in]{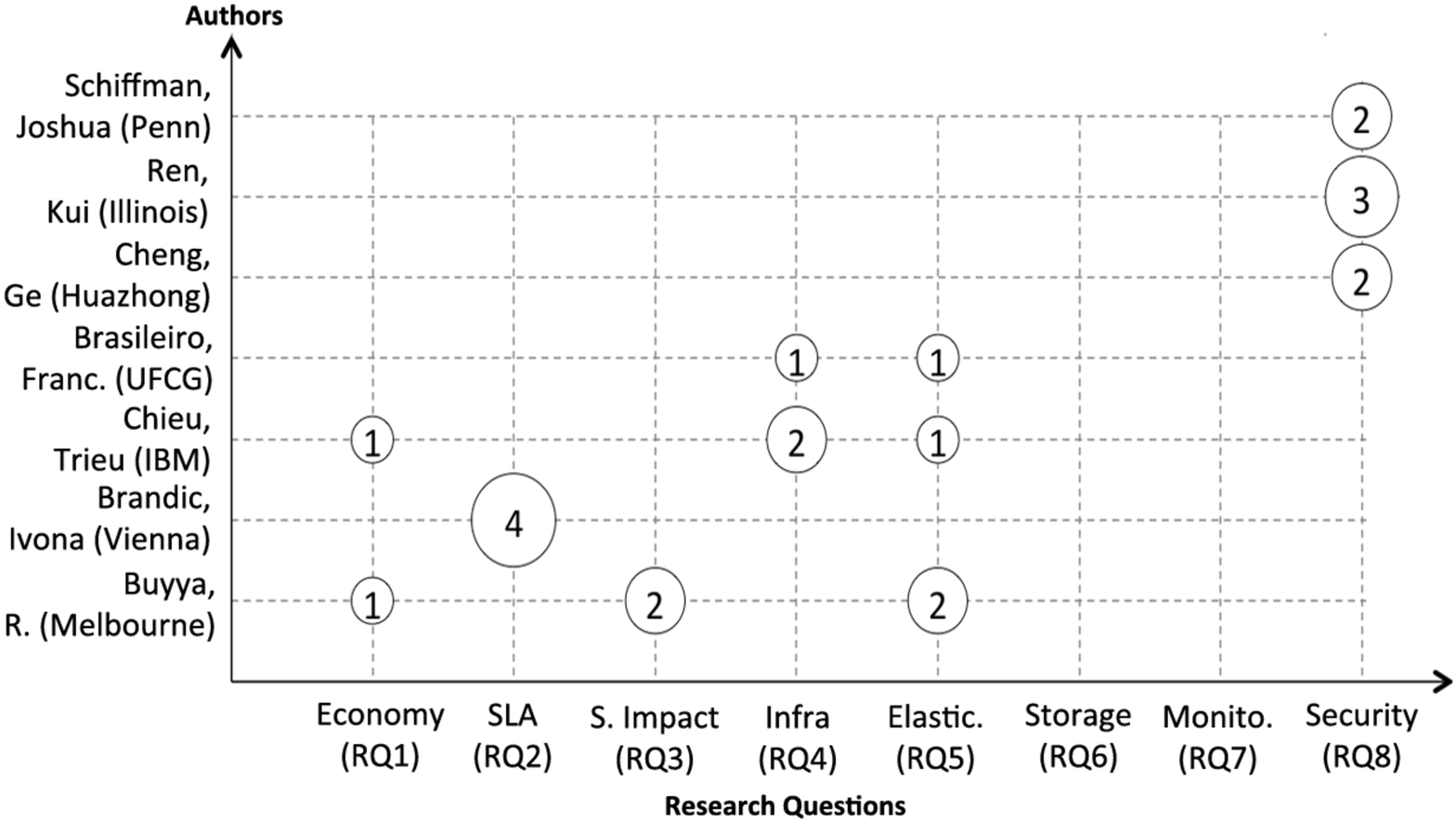}
\caption{Authors/Institutes by Research Focus}
\label{fig_authorsbyrq}
\end{figure}

Also, a list of nontrivial field were found in this research and presented on Appendix C.
\section{Discussion}
For Ichak Adizes \cite{fish:IchakAdizes}, the main factors for the death of enterprises during the first 5 years (Infancy) are:
\begin{itemize}
\item lack of planning;
\item uncertainties about the customers, suppliers and market;
\end{itemize}

In this context, the evidences points that Cloud Computing can fit as an element that allows the control of expenses. This is possible thanks to its proposal on-demand. However, there remain some caveats. 

Parallel applications has being used not only by scientific  projects but also by organizations and Government. However, the studies points that Cloud Computing has presented limitations in that application class. This occurs because of limitations both resource provisioning and business model of the Cloud Providers. Therefore, it is understood that Cloud is not yet suitable for the market \cite{fish:Kondo}  \cite{fish:Sevior}.

Another point is the complexity for implementation of a Cloud environment. Nae et al. \cite{fish:Nae} affirm to have obtained a better efficiency level with a \textbf{Private Cloud} model than conventional data center model. However, the authors claim that the complexity for implement the model was a challenge.


 \hvFloat[
 floatPos=hbp,
 capWidth=h,
 capPos=r,
 capAngle=90,
 objectAngle=90,
 capVPos=c,
 objectPos=c]{figure}{ \includegraphics[width=7in,height=4.3in]{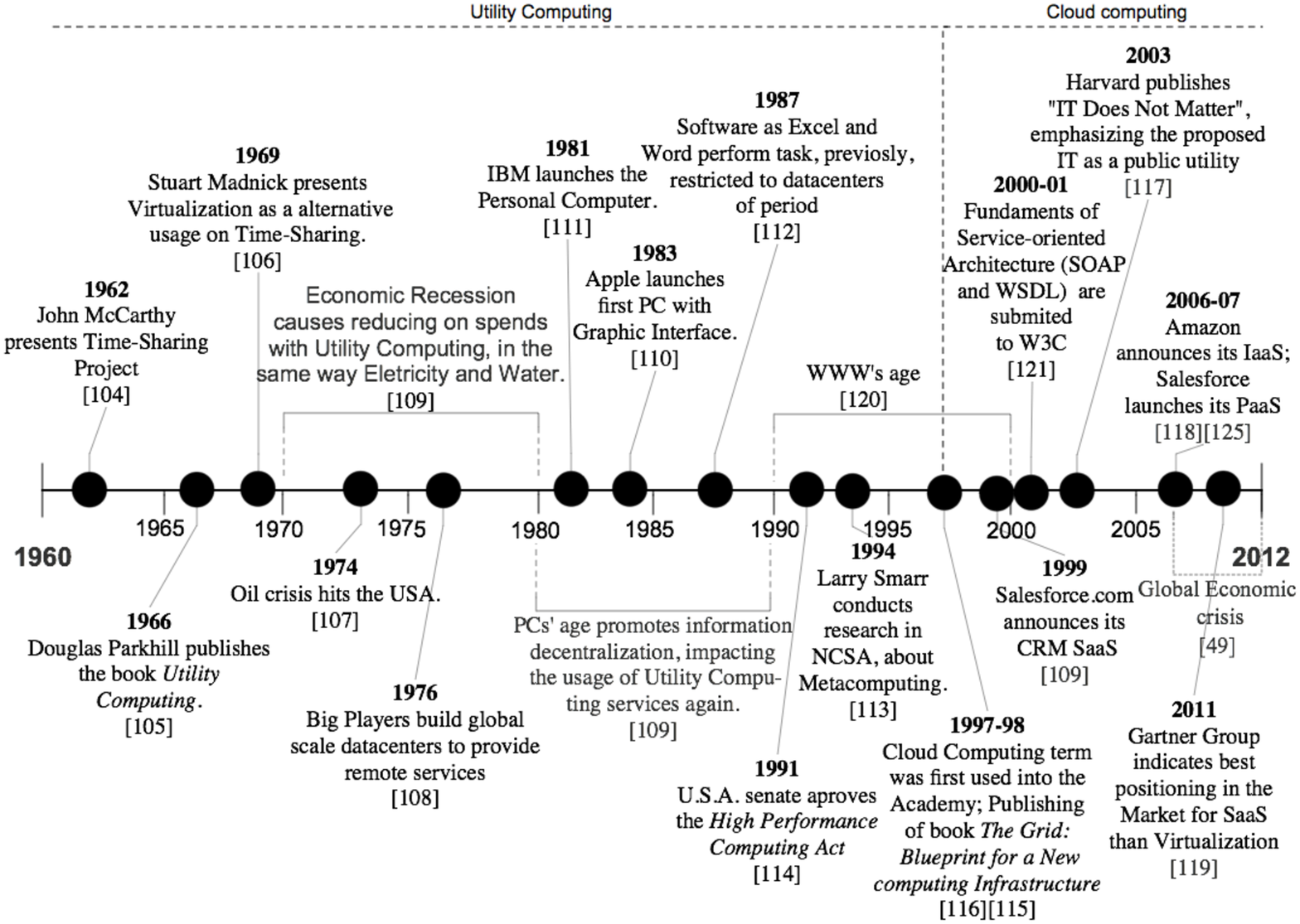} \hspace*{0.7in}}%
{Cloud Computing Timeline.}{fig_timeline}

Therefore, in adoption process, we suggest be careful with the upfront low costs promise. Including, were waited for more evidences about costs of migrations from conventional datacenters to Cloud environments in order to compare scenarios. However, it is not possible.

A good structured SLA is the better way to safeguard the two parts, Cloud Provider and Customer \cite{fish:Luo}\cite{fish:Chaves}. Then, we suggest that before begin a migration process to Cloud or to provide a service as a \textbf{Public Cloud}, the two parts need a good structured SLA contract. Otherwise, the stakeholders may undergo serious problems of security, loss of privacy and other conflicts \cite{fish:Kandukuri}.

For a future where Cloud Computing will be a green IT approach, we suggest more detailed researches as shown by Baliga et. al  \cite{fish:Baliga}. In that work, the author treats the problem of energy consumption beyond the limits of the datacenter.

Some studies found in this mapping \cite{fish:ChoBrian} \cite{fish:Taylor} \cite{fish:Doelitzscher} \cite{fish:Udo} \cite{fish:Zhou} shown that the conflicts between Cloud and contemporary laws inspire new shifts. This is because of scenarios as at Federal Data Protection Act \cite{fish:ChoBrian},  section 11. In that, it is described that a customer must know the location of its data. But, Cloud Computing can put a data on anywhere because its nature.

However, many factors should be studied about that. Thus, we suggest that Cloud Providers pay attention at Laws before distribute its Cloud environments among countries.

Based on evidences, it was possible to create a timeline about Cloud, confirming the evolution from Utility Computing to Cloud Computing, as shown on Figure \ref{fig_timeline}. Also, by analyzing of Figure \ref{fig_timeline}, it was possible identify that the 70's reported a reduction in the use of Utility Computing due to recession of that period. However, nowadays, even with an international crisis scenario, the use of Cloud Computing is increasing. This is because computing limitation that period, setting limits for creation of new services models. 

It is important to mention that at that time was common to find computers contained Memory RAM with 1K \footnote{http://oldcomputers.net/vector1.html}.

Still is unclear the factors that must be into account, when building a system architecture to provide service on Cloud Computing\cite{fish:Kossmann}. However, based on studies is understood that a suitable development of a XaaS must be composed by harmony between its infrastructure and software application, transforming a heterogeneous environment to homogeneous. This way, allowing to the service to be delivered in a flexible way (on-demand) and meeting requirements of all customer classes.

For conception of a SaaS, we found, among other aspects, as follow: model-driven style \cite{fish:Menzel}, multi-tenant maturity levels \cite{fish:Cadan}, Service-Oriented Architecture \cite{fish:Hutchinson} and RESTFull model \cite{fish:Calero}. But, to know which techniques or the best way to use them, we suggest a search more deep into context of architectures of Cloud. This is because that specific issue is not the focus of this work.

The search by Science \cite{fish:Hou_Zhengxiong} \cite{fish:Vecchiola} \cite{fish:Lu Wei} \cite{fish:He Qiming} \cite{fish:Ramakrishman} for reducing of cost on projects also is an important fact for future investments on Cloud. However, the stake-holders must know that Science groups have specific requirements such as payment methods for services \cite{fish:Ramakrishman} or parallel processing \cite{fish:Lu Wei}.

However, we also agree with idea that on Grid and Supercomputer context, the systems are based on optimizing peak floating-point performance for individual applications per megawatt of energy \cite{fish:Hassan}. On the other hand, Cloud systems must be based on user experience per megawatt by hour.

The interoperability among Public Clouds can be a interesting theme to future researches and the access control is one of solutions \cite{fish:Calero} \cite{fish:Li Jin}. However, Cloud Computing needs of Open Standards such as \textbf{Open Virtualization Format (OVF)} and \textbf{Security Assertion Markup Language (SAML)} of OASIS to reach other horizons.

\section{Threats of Validity}
There are some threats to the validity of our study, which we briefly describe along with the mitigation strategy for each threat:
\begin{itemize}
\item	Research Questions: The set of questions that we defined might not have covered the whole Cloud Computing area, which implies that one cannot find answers to the questions that concern them. As we considered this as a feasible threat, we had several discussion meetings with project members, SMaRT research group members in order to calibrate the questions. In this way, even if we had not selected the most optimum set of questions, we attempted to address the most asked and considered open issues in the field.
\item	Publication Bias: We cannot guarantee that all relevant primary studies were selected. It is possible that some relevant studies were not chosen throughout the searching process. We mitigated this threat to the extent possible by following references in the primary studies. 
\item	Conduct the Search: The digital databases do not have a compatible search rules. We adapted our search strings for each digital database. However, it is important mention that ACM digital library presented problem in its search engine on February of 2011. With this we spent a lot of time to find a way of reach correct results in this engine.
\item	Data Extraction: During the extraction process, the studies were classified based on our judgment. However, despite double-checking, some studies could have been classified incorrectly. In order to mitigate this threat, all process of classification was showed for SMaRT research group members.
\end{itemize}

\section{Conclusion}
In this study was presented a systematic mapping study on Cloud Computing. Through that method we conducted this research investigating the state-of-the-art in Cloud Computing, clarifying open issues through a analysis of evidences found in 301 primary studies.

Through the answers found in eight research questions, it was possible to identify evidences that point to Cloud Computing as an emerging approach, which proposes a shift of paradigm in the context of Information Technology, enabling a rational model of computation, based on Utility Computing.

In general, Cloud Computing still needs improvements that enable the heterogeneity of its elements work in harmony, in order to transform the current model of computing a truly on-demand environment. However, studies like \cite{fish:Nae} show concretely the efficiency of the model. However, we waited for more evidences about that subject.

Based on evidences, we build a Cloud Computing timeline from 60`s to 2012. That way, it was possible identify market behaviors over the last 50 years and the closed relation between Utility Computing and Cloud Computing models. Thus, characterizing Cloud Computing as a feasible approach to nowadays, considering the past facts.

Security and Privacy are big issues related to adoption process of Cloud solutions \cite{fish:Popovic} \cite{fish:Yu Xiaojun}. However, through this study it was possible to identify a big and continuous effort of the community in order to mitigates the problems around  these topics. Thus, in the adoption process, we understand that the interference of Security and Privacy issues will be minimized over course of time.

This study also identified that Cloud Computing requires efforts regarding the deployment of parallel applications in your environment. Although we find initiatives around the subject, these are early stage.

Another identified problem was the inadequacy of business models. That way, forming barriers to the adoption of Cloud Computing solutions. Because, if a Cloud Provider makes its services available based on credit card transactions, how can a science group to introduce itself in this environment?.

We understand that the services diversity proposed by Cloud Computing (XaaS) is one of the most attractive elements of the model. Thus, based on evidences, we enumerate some suggestions about services on Cloud:
\begin{itemize}
\item	make sure that the service is really under a Cloud environment. Thus, promoting the use of benefits such as flexibility, time to market, pay-per-use and ubiquity;
\item the whole Cloud environment  must be implemented focusing on green IT and dealing with power energy factors beyond the datacenters \cite{fish:Baliga};
\item by distributing the datacenters among countries, the Cloud Provider must pay attention to local laws. Thus, eliminating potential complications;
\item to adapt the service model and Cloud environment according to the customer.
\end{itemize}

The results achieved by this mapping study will help our research group to develop new research fronts about Cloud Computing.  Our next step, will be to conduct a research toward the impacts of Cloud Computing in Software Engineering. 

\section{Acknowledgment}
The author would like to thank SMaRT Research Group for the feedback on meetings and support.

%
\newpage
\section{Appendix A}
This section presents the lists of 10 first conferences (Table \ref{table_conferences}) and journals (Table \ref{table_journals}) ordered by the number of selected primary studies. 

The whole list of sources are available at the public link: http://goo.gl/O7jcS.
\begin{table}[hbp]
\renewcommand{\arraystretch}{1.3}
\caption{List of 10 first conferences ordered by the number of selected primary studies.}
\label{table_conferences}
\centering
\begin{tabular}{| p{9cm} | p{1cm}|}
\hline
\bfseries Event & \bfseries N.\\
\hline
International Conference Cloud Computing Technology and Science (CloudCom) & 11\\
\hline 
International Conference Cloud Computing (CLOUD) & 10\\
\hline
IEEE/ACM International Symposium on Cluster, Cloud and Grid Computing (CCGrid) & 6\\
\hline
World Congress on Services (SERVICES-I) & 4\\
\hline
IEEE/IFIP Network Operations and Management Symposium (NOMS) & 4\\
\hline
IEEE International Conference on E-Business Engineering (ICEBE) & 4\\
\hline
ACM Symposium on Cloud Computing (SoCC) & 4\\
\hline
ACM Conference on Computer and Communications Security (CCS) & 4\\
\hline
International Conference on Grid and Cooperative Computing (GCC) & 3\\
\hline
International Conference on Computer and Information Technology (CIT) & 3\\
\hline
\end{tabular}
\end{table}

\begin{table}[hbp]
\renewcommand{\arraystretch}{1.3}
\caption{List of 10 first journals ordered by the number of selected primary studies.}
\label{table_journals}
\centering
\begin{tabular}{| p{9cm} | p{1.0cm}|}
\hline
\bfseries Title & \bfseries N.\\
\hline
IEEE Security \& Privacy & 8\\
\hline 
The Journal of Supercomputing & 5\\
\hline
Communications of the ACM& 4\\
\hline
Grid Computing & 4\\
\hline
IT Professional & 4\\
\hline
Annals of telecommunications & 3\\
\hline
Computer Law \& Security Review & 2\\
\hline
Computing in Science \& Engineering & 2\\
\hline
Future Generation Computer Systems & 2\\
\hline
IEEE Transactions on Parallel and Distributed Systems & 2\\
\hline
\end{tabular}
\end{table}

\newpage
\section*{Appendix B}
It is important mention that studies referenced \textbf{only} in Table \ref{table_references} are available at the public link: http://goo.gl/A29DY.

The references in bold represent the most relevant studies due to its clarity and fullness on the issues addressed in this research.
\begin{table}[hbp]
\renewcommand{\arraystretch}{1.3}
\caption{Classification of References}
\label{table_references}
\centering
\begin{tabular}{| p{1cm} | p{10cm}|}
\hline
\bfseries RQs & \bfseries Studies\\
\hline
RQ1 & \textbf{[5]}, \textbf{[7]}, [36], [49], \textbf{[50]}, \textbf{[51]}, [52], \textbf{[53]}, [30], [94], \textbf{[91]}, [145], [146], [147], [148], [149], \textbf{[109]}, [150], [151], [152], [153], \textbf{[303]}, [97], [100].
\\
\hline 
RQ2 & \textbf{[2]}, \textbf{[3]}, \textbf{[54]}, \textbf{[4]}, [31], [35], [155], [156], [157], [158], [122], [159], [160], [161], [123], [162], [163], [164], [165], [166], [167], [168].
\\
\hline
RQ3 &\textbf{[9]}, \textbf{[10]}, \textbf{[11]}, \textbf{[303]}, \textbf{[13]}, \textbf{[56]}, [14], [32], [33], [169], [170], [171], [172], [173], [174], [175], [176], [177], [178], [179], [180], [181], [182], [183], [184], \textbf{[185]}, [186], [187], \textbf{[188]}, [189], [190], [191], [192], \textbf{[193]}, \textbf{[194]}, \textbf{[195]}, \textbf{[196]}, \textbf{[197]}.\\
\hline

RQ4 &[57], \textbf{[1]}, \textbf{[60]}, \textbf{[124]}, [59], \textbf{[6]}, [15], [198], [87], \textbf{[88]}, [89], [90], \textbf{[93]}, [191], [200], [99], [201], [202], [203], [204], [205], [206], [207], [208], [209], [210], [211], [212], [213], [214], [215], [216], [217], [218], [219], [220], \textbf{[221]}, [140].\\
\hline
RQ5 &\textbf{[8]}, \textbf{[303]}, \textbf{[62]}, \textbf{[66]}, \textbf{[65]}, \textbf{[222]},[64], [67], \textbf{[223]}, [42], [224], [225], [226], [227], [228], [229], [230], [231], [232], [233], [234], [143], [235], [236], [237], [238], [239], [240], [241], [242], [243], [244], [245], [246], [247], [248], [249], [250], [251], [252], [253], [254], [255], [256], [257], [258], [259], [260], [98], \textbf{[261]}, [342].
\\
\hline
RQ6 &  [68], [69], [74], [96], [92], [70], [71], [75], [73], [78], [49], [80], [72], [262], [263], [264], \textbf{[265]}, \textbf{[266]}, [267], [268], [269], [270], [271], [272], [273], [274], [275], [144], [276], [277], [278], [45].
\\
\hline
RQ7  & [79], [58], [126], [279], [127], [129], [280], [281], \textbf{[128]}.
\\
\hline
RQ8 & \textbf{[23]}, \textbf{[24]}, [85], \textbf{[25]}, [86], [26], [27], [81], [82], \textbf{[28]}, [84], [282], \textbf{[29]}, [38], [95], \textbf{[44]}, \textbf{[283]}, [284], [133], \textbf{[285]}, [286], [287], \textbf{[288]}, [289], [290], [291], [292], [293], \textbf{[141]}, [294], [295], [296], [134], [135], [297], [298], [299], [300], [301], [136], [130], [302], \textbf{[303]}, [304], [305], [306], [307], [308], [309], [310], [311], \textbf{[312]}, [313], [314], [315], [316], [317], [318], [319], [320], [321], [322], [323], [324], [325], [326], \textbf{[327]}, [328], [329], [330], [331], [132], [332], [333], [131], [334], [335], [336], [138], [337], \textbf{[142]}, [338], [339], [139], [340], [137], [341].
\\
\hline
\end{tabular}
\end{table}

\newpage
\section{Appendix C}
The Table \ref{table_fields} presents nontrivial areas working with Cloud Computing. Thus, presenting opportunities of more deep studies.

\begin{table}[hbp]
\renewcommand{\arraystretch}{1.3}
\caption{Nontrivial fields working with Cloud.}
\label{table_fields}
\centering
\begin{tabular}{| p{4cm} | p{7cm}|}
\hline
\bfseries Field & \bfseries Description\\
\hline
Health Care & Medical images, storing patient data, privacy \cite{fish:Chia Ping} \cite{fish:Yang}.\\
\hline 
e-Learning & Sharing data among students \cite{fish:Ma Hui}.\\
\hline

GeoSpatial Computing & Uses GPU (graphical process Unit) on servers \cite{fish:Zhang Jianting}.\\
\hline
Textil Industry (China) & Setting up ERP on Cloud \cite{fish:Zhang Kejing}.\\
\hline
NASA (Jet Propulsion) & Storage \cite{fish:Mattmann Chris}.\\
\hline
e-Government and e-Voting & A way to improve services for citizens \cite{fish:Zissis}.\\
\hline
China Railway & Scalibitiy for Services on system \cite{fish:Baopeng}.\\
\hline
Chemical and Petroleum & Infraestructure use \cite{fish:Zhou Yu}.\\
\hline
\end{tabular}
\end{table}

\newpage

\end{document}